\documentclass[jog]{igs}

\usepackage{appendix}
\usepackage[utf8]{inputenc}
\usepackage{float}
\usepackage{dblfloatfix}
\usepackage{placeins}
\usepackage{graphicx}
\usepackage{siunitx}
\usepackage{amsmath}
\def\RIP{$n(z)$~}

\def\nX#1{\textit{n}_{\text{#1}}(z)}

\begin{document}

\title[Probing the Firn Refractive Index Profile Using Antenna Response]{Probing the Firn Refractive Index Profile Using Antenna Response}

\author[Agarwal and others]{
  S. Agarwal$^1$,
J.~A. Aguilar$^2$,
N. Alden$^{3}$,
S. Ali$^1$,
P. Allison$^4$,
M. Betts$^5$,
D. Besson$^{\star, 1}$,
A. Bishop$^{6}$,
O. Botner$^7$,
S. Bouma$^8$,
S. Buitink$^{9,10}$,
R. Camphyn$^{2}$,
S. Chiche$^{2}$,
B.~A. Clark$^{11}$,
A. Coleman$^{7}$,
K. Couberly$^1$,
S. de~Kockere$^{12}$,
K.~D. de~Vries$^{12}$,
C. Deaconu$^{3}$,
P. Giri$^{13}$,
C. Glaser$^7$,
T. Gl{\"u}senkamp$^7$,
A. Hallgren$^7$,
S. Hallmann$^{14,8}$,
J.~C. Hanson$^{15}$,
B. Hendricks$^5$,
J. Henrichs$^{14,8}$,
N. Heyer$^7$,
C. Hornhuber$^1$,
E.~Huesca Santiago$^{14}$,
K. Hughes$^4$,
T. Karg$^{14}$,
A. Karle$^6$,
J.~L. Kelley$^6$,
M. Korntheuer$^{2,12}$,
M. Kowalski$^{14,16}$,
I. Kravchenko$^{13}$,
R. Krebs$^5$,
R. Lahmann$^8$,
C.-H. Liu$^{13}$,
M.~J. Marsee$^{17}$,
C. McLennan\thanks{Corresponding Author e-mail: curtis.mclennan@ku.edu, zedlam@ku.edu, authors@rno-g.org}$^{\star,1}$,
M. Mikhailova$^1$,
K. Mulrey$^{10}$,
M. Muzio$^6$,
A. Nelles$^{14,8}$,
A. Novikov$^{18}$,
A. Nozdrina$^4$,
E. Oberla$^{3}$,
B. Oeyen$^{19}$,
N. Punsuebsay$^{18}$,
L. Pyras$^{14,8}$,
M. Ravn$^7$,
D. Ryckbosch$^{19}$,
F. Schl{\"u}ter$^2$,
O. Scholten$^{12,20}$,
D. Seckel$^{18}$,
M.~F.~H. Seikh$^1$,
J. Stachurska$^{19}$,
J. Stoffels$^{12}$,
S. Toscano$^2$,
D. Tosi$^6$,
J. Tutt$^{5}$,
D.~J. Van~Den~Broeck$^{12,9}$,
N. van~Eijndhoven$^{12}$,
A.~G. Vieregg$^{3}$,
A. Vijai$^{11}$,
C. Welling$^{3}$,
D.~R. Williams$^{17}$,
P. Windischhofer$^{3}$,
S. Wissel$^5$,
R. Young$^1$,
A. Zink$^8$
}
\affiliation{
$^1$University of Kansas, Dept.\ of Physics and Astronomy, Lawrence, KS 66045, USA\\
$^2$Universit\'e Libre de Bruxelles, Science Faculty CP230, B-1050 Brussels, Belgium\\
$^{3}$Dept.\ of Physics, Dept.\ of Astronomy \& Astrophysics, Enrico Fermi Inst., Kavli Inst.\ for Cosmological Physics, University of Chicago, Chicago, IL 60637, USA\\
$^4$Dept.\ of Physics, Center for Cosmology and AstroParticle Physics, Ohio State University, Columbus, OH 43210, USA\\
$^5$Dept.\ of Physics, Dept.\ of Astronomy \& Astrophysics, Center for Multimessenger Astrophysics, Institute of Gravitation and the Cosmos, Pennsylvania State University, University Park, PA 16802, USA\\
$^6$Wisconsin IceCube Particle Astrophysics Center (WIPAC) and Dept.\ of Physics, University of Wisconsin-Madison, Madison, WI 53703, USA\\
$^7$Uppsala University, Dept.\ of Physics and Astronomy, Uppsala, SE-752 37, Sweden\\
$^8$Erlangen Centre for Astroparticle Physics (ECAP), Friedrich-Alexander-University Erlangen-N\"urnberg, 91058 Erlangen, Germany\\
$^9$Vrije Universiteit Brussel, Astrophysical Institute, Pleinlaan 2, 1050 Brussels, Belgium\\
$^{10}$Dept.\ of Astrophysics/IMAPP, Radboud University, PO Box 9010, 6500 GL, The Netherlands\\
$^{11}$Department of Physics, University of Maryland, College Park, MD 20742, USA\\
$^{12}$Vrije Universiteit Brussel, Dienst ELEM, B-1050 Brussels, Belgium\\
$^{13}$Dept.\ of Physics and Astronomy, Univ.\ of Nebraska-Lincoln, NE, 68588, USA\\
$^{14}$Deutsches Elektronen-Synchrotron DESY, Platanenallee 6, 15738 Zeuthen, Germany\\
$^{15}$Whittier College, Whittier, CA 90602, USA\\
$^{16}$Institut f\"ur Physik, Humboldt-Universit\"at zu Berlin, 12489 Berlin, Germany\\
$^{17}$Dept.\ of Physics and Astronomy, University of Alabama, Tuscaloosa, AL 35487, USA\\
$^{18}$Dept.\ of Physics and Astronomy, University of Delaware, Newark, DE 19716, USA\\
$^{19}$Ghent University, Dept.\ of Physics and Astronomy, B-9000 Gent, Belgium\\
$^{20}$Kapteyn Institute, University of Groningen, PO Box 800, 9700 AV, The Netherlands\\

}

\begin{frontmatter}

\maketitle

\begin{abstract}
The Radio Neutrino Observatory-Greenland (RNO-G, at Summit Station) experiment comprises an extensive fat-dipole antenna array deployed into ice boreholes over an eventual area of approximately 35 ${\rm km}^2$. Since the RNO-G experimental sensitivity depends on the radio-frequency properties of the firn, which are known to vary laterally on sub-km distance scales and vertically on sub-meter distance scales, a technique for quickly extracting information on firn ice properties with depth ($n(z)$) during drilling and deployment is desirable. Given that 
a dipole's resonant wavelength is fixed by geometry, the resonant frequency $f_{res}$ (measured as an S-parameter reflection coefficient [`$S_{11}$'] minimum)
scales inversely with the local refractive index, allowing a translation of
 a depth-dependent $S_{11}$(z) profile into $n(z)$. $S_{11}$(z) data were initially taken in August, 2024 using a dipole lowered into a newly-drilled $98\pm 1$-mm diameter, 350-m deep borehole at Summit Station, Greenland, approximately 1 km from the site of the original GISP-2 core; improved measurements were subsequently made in May, 2025. We conclude that $S_{11}$(z) data can be used to estimate \RIP, on 50 cm vertical scales, at the per-cent level of accuracy required by experiments such as RN0-G.

\end{abstract}

\end{frontmatter}

\section{Introduction}
\subsection{Radio-frequency neutrino detection in cold polar ice}
\begin{figure}[h]
     \centering
     \includegraphics[width=\linewidth]{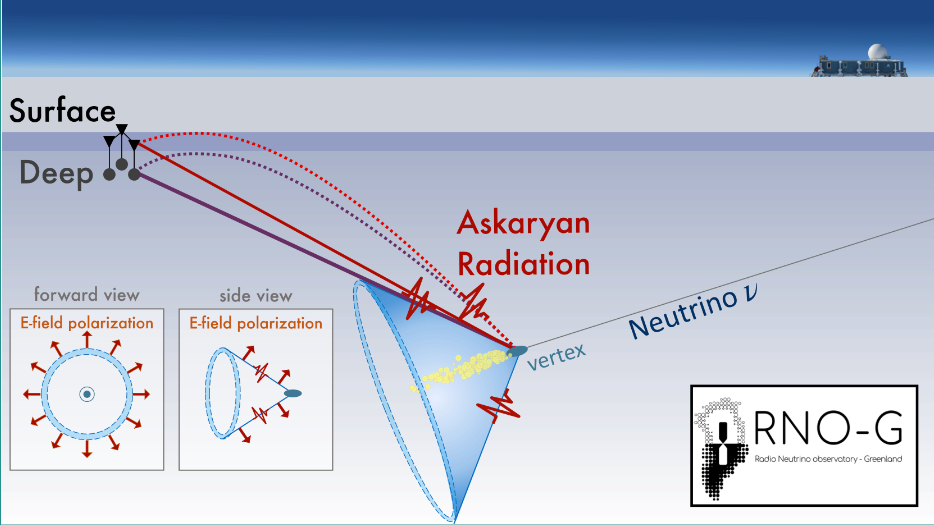}
     \caption{Schematic illustrating UHEN signal detection. An incident neutrino collides with an ice molecule at a `vertex', producing Askaryan radiation (consisting of radially outward polarized propagating electric fields) geometrically confined to the surface of an expanding Cherenkov cone. As a result of the variable refractive index in the firn (deep purple), two rays (direct [`D'; solid lines] and refracted/reflected  ['R'; dashed lines]) reach each of the Surface and also Deep receiver antennas. Purple lines show the highest-amplitude, `on-Cone' signal; Red lines show weaker, `off-Cone' signal.}
     \label{fig:pictogram}
 \end{figure}

Over the last three decades, aided by significant advances in digital technology, detection of  coherent 
radio-frequency (RF) Cherenkov-like radiation \citep{Askaryan:1961,Askaryan1965CoherentRE}, or RF echoes from neutrino-induced cascades \citep{prohira2021radar} in dense media have emerged as the most cost-effective approaches to measure Ultra-High Energy Neutrinos (UHEN). Such UHEN detectors need to monitor $\mathcal{O}(\qty{10}{\km^3})$ volumes to be sensitive to the extremely low UHEN flux at neutrino energies $\sim$ $\qty{100}{\peta \eV}$ and above. 
Abundant, high purity polar ice provides a nearly ideal dense target medium, featuring $\mathcal{O}(\qty{1}{\km})$ RF attenuation lengths \citep{robin1969interpretation,barwick2005south,aguilar2022situ}, and typically moderate anthropogenic noise environments. Fig. \ref{fig:pictogram} illustrates the neutrino detection scheme; further reviews of UHEN detection using RF antennas can be found in the literature \citep{huege2017radio,Schr_der_2017}.

\subsection{Radio-frequency ice permittivity and the impact on UHEN experiments}
Accurate and precise characterization of the RF response of glacial firn is essential for UHEN experiments seeking observation of neutrinos using radio techniques, as the dielectric permittivity in the hundreds of MHz frequency regime directly impacts the active volume over which neutrino interactions can be measured (`effective volume') and also the ability to point an interacting neutrino back into the sky (`neutrino astronomy').  
One parameter of particular interest is the depth-dependent refractive index profile (`RIP', or $n(z)$), which determines the speed of light in different media and therefore dictates signal trajectories. For RNO-G, with antennas deployed 1--20 meters apart vertically, and 1.25 km laterally between receiver stations, 1\% precision in local RIP measurements, is required in order that \RIP uncertainties not dominate neutrino-finding and reconstruction errors.

\subsection{Refractive Index (RI) Modeling}
The dielectric response of a given medium to an electromagnetic excitation of angular frequency $\omega$ is often formulated in the Drude-Lorentz atomic-resonator model \citep{drude1900elektronentheorie,lorentz1916theory}, for which the equation of motion is calculated assuming an incident vector Electric field ${\vec E}$ driving an atomic electron, modulo damping (via interatomic effects and giving rise to extinction) and resonance effects. Ice lattice interactions produce an absorptive resonance pole in the complex pure ice permittivity at far-infrared frequencies (4.8 and 14.48 THz). At radio frequencies and below, and extending into the hundreds of kHz regime, ice response is characterized by temperature-dependent Debye relaxation, corresponding to small absorptive losses and therefore long attenuation lengths. 

The Fujita et al. compilation of extant complex dielectric data \citep{fujita2000summary} extensively summarizes both lab and {\it in situ} measurements of the complex ice permittivity, indicating a $\sim$0.5\% systematic uncertainty in the refractive index, with a 0.2\% variation over the temperature range relevant for UHEN experiments. Consistent with expectations from the atomic resonator model, the dependence of the complex permittivity on wavelength over the radio-frequency band is mild. In their review, the authors specifically consider the physical attributes of natural ice (density, air bubble infusion, impurity [particularly conducting acids and free $H^+$ ions] and temperature) affecting the refractive indices, finding density to be the dominant factor, and supporting an assumed linear dependence of RI on density. Their review of the literature indicates a secondary role played by temperature; translating their compiled data into RI temperature dependence, their data imply $n(T)$=$n(T_0$)+$\alpha(T-T_0)$, with $n(T_0=190$K)=1.77 and $\alpha=8.3\times 10^{-5}$ over our relevant temperature range. There was no conclusive frequency dependence (dispersion) indicated by current experimental results; in fact, strong limits have been placed at both South Pole and also Summit Station, ruling out dispersion as an important contributor to UHEN experimental systematics \citep{kravchenko2011radio,aguilar2023radiofrequency}.

\subsection{RI Measurements to-date}
The real part of the permittivity ($\varepsilon_r$) can be directly quantified in the lab via dielectric profiling (DEP) of ice cores, the small lateral size of the cores being profiled, relative to radio wavelengths, notwithstanding.

Gravimetric density measurements are also straightforward, and can be translated to permittivity assuming a one-parameter dependence \citep{schytt1958snow,KOVACS1995245}:

\begin{equation}
\varepsilon_r = \Bigg[1+0.845\rho\Big(\frac{\unit{\g}}{\unit{\cm^3}}\Big)\Bigg]^2;~~ n=\sqrt{\varepsilon_r}.
\end{equation}

Neutron probe monitor (NPM) measurements \citep{morris2003density,Hawley_Morris_2006}, made in an ice borehole for which the hole diameter has already been measured to a precision of 400 microns can, in principle, measure 2\% density (statistical) variations on a vertical length scale of 1 cm if neutron back-scatter can be measured at each sampling point for 64 seconds (the hole diameter must be known in order to precisely calibrate out residual back-scatter unassociated with the ice volume itself).

The bulk index-of-refraction can also be derived {\it in situ} by the UHEN experiments using transit time (TT) measurements of polarized impulsive (ns-duration) signals propagating over a macroscopic distance {\tt d}, for which TT={\tt d}$n/c_0$. This technique was recently applied to surface-based radar-echo sounding (RES) data taken at Summit Station \citep{welling2024brief}, for which the time difference between successive internal layer echoes could be determined to nanosecond-scale accuracy. Correlating the echo times with decimeter-scale tabulations of internal layer depths from archived ice cores in the vicinity of Summit Station provided the most precise determination, to date, of the bulk polar ice refractive index ($n$=1.778$\pm$0.006) at depths below several hundred meters \citep{welling2024brief}.

\subsection{RNO-G RIP parameterization and current limitations}
Translations of density into \RIP are one important component of the RIP used by the RNO-G experiment \citep{RNOG_Philipp}.
Performing a fit over all the available techniques and data yields the RIP favored by the RNO-G experiment \citep{agarwal2025instrument,RNOG_Philipp}.
One major drawback of these fits, however, is that they yield a single (smooth) \RIP parameterization for the entire 35 ${\rm km}^2$ RNO-G footprint. However, significant (5--10\%) local variations in $n(z)$ have been already noted at sites less than 1 km away from each other \citep{morris2003density,Hawley_Morris_2006} near Summit.

One-parameter $n(z)$  formulations are also insensitive to the (known) 
tensorial characteristics of the dielectric permittivity. Several experiments have demonstrated polarization-dependent anisotropies in RF propagation depending on the orientation of the electric field vector relative to the ice crystal orientation fabric.  
As is well-known at optical frequencies, and also now well-measured at radio frequencies\\\citep{kravchenko2011radio,allison2019measurement,besson2023polarization}, birefringence results in a wavespeed asymmetry for polarizations parallel vs. perpendicular to the crystal orientation fabric. Fujita {\it et~ al.} predict a 0.5\% asymmetry, although direct {\it in situ} experiments have only measured (at most) half that value, perhaps as a result of incomplete alignment of the crystal orientation fabric (COF).

Additionally, (known) density fluctuations can disrupt the coherence between the frequency components that comprise the sought-after otherwise-coherent Cherenkov radiation signal electric field vector, for either neutrinos observed via Refracted signals with trajectories sampling the near-surface ice, or down-coming cosmic-ray-induced showers with steep inclination angles. Recently, the Askaryan Radio Array claimed first-ever observation of such down-coming cosmic ray signals \citep{alden2025observation}. 
Numerically, we can estimate the magnitude of decoherence from density fluctuations, knowing that the number of particles contributing to the coherent Cherenkov signal is approximately the shower energy, in GeV, divided by 4. A shower having energy $10^{17}$ eV, or $10^8$ GeV, would therefore be reduced in amplitude by a factor of 10,000 in the extreme case of full coherence loss, underscoring the importance of mapping any density fluctuations on a length scale comparable to the smallest measurable wavelengths (20 cm) for the UHEN experiments.

Overdense melt layers within the firn, for example, can vary in thickness from centimeter to meter scale, with local density variations that can approach that of ice, and therefore produce significant coherence loss \citep{nghiem2012extreme}. Quantifying the magnitude of density fluctuations (which are empirically observed to decrease in magnitude with increasing depth) can (hopefully) allow us to determine the impact on UHEN and UHECR astronomy.\footnote{Determining the numerical implications of density fluctuations on measurements of Askaryan radiation from impacting air shower cores is a primary goal of an upcoming Summit Station field season.}

\subsection{Glacial Firn Ice Properties}
\label{sec:Ice}
The antennas of the RNO-G experiment are deployed at maximum depths of 100 m, within the firn layer of the ice sheet.
Firn properties change with depth, as the ice gradually compacts with increasing overburden. 
Unlike the Earth’s atmosphere, which readily lends itself to a simple exponential density profile for a fluid compressing under
its own weight, compaction of accumulated snow proceeds through multiple density transitions, often
demarcated by two separate inflection points \citep{Herron1980FirnDA,gmd-13-4355-2020,Salamatin_Lipenkov_Duval_1997}. At a density of 550 ${\rm kg/m^3}$, corresponding to a depth of 12 m at Summit Station, snow (depending on moisture content) crystals
transform to ice characterized by an abundance of grains and a density intermediate between snow
and glacial ice. At a density of 830 ${\rm kg/m^3}$, `close-off' occurs and bubbles can no longer mix with surface air. At this depth ($\sim$75--80 m at Summit), firn ice transforms to ‘bubbly ice’, beyond which the ice structure
is relatively constant, modulo a diminishing fraction of air inclusions. Over the upper 100 m, the density
profile thus evolves from $\sim$45\% of the asymptotic deep ice density (917 kg/${\rm m^{3}}$) to nearly 
$90-95\%$ of the asymptotic value. This density change corresponds to a refractive index range of $\sim$1.3–1.75 at Summit; averaged over the entire Greenland Ice Sheet, the average surface (upper 10 cm) ice density has been measured to be 315$\pm$44 kg/${\rm m^{-3}}$ \citep{fausto2018snow}, corresponding to a surface RI value $\nX{surface}=1.27$. 

Thus far, numerous models have been advanced to parameterize firn densification relative to the surface; each\\ model should ideally include a quantitative description of multiple factors contributing to the observed $\rho(z)$ density profile, including local moisture content, local snowfall rate, temperature profile and heat diffusion through the ice, diffusion of both water and above-surface air, grain growth and evolution of the local ice fabric, etc. The large number of input priors underscores the importance of obtaining {\it in situ} density data \citep{vandecrux2020firn}. More extensive information regarding ice, firn and scattering and densification processes can be found in the literature \citep{Herron1980FirnDA,fujita2000summary,cuffey2010physics,horhold2011densification}. 

\section{Antenna Response In Media and sensitivity to RIP}
Antennas emit/receive electromagnetic radiation, converting the emitted/captured energy from/to electrical currents. The frequency bandwidth over which an antenna responds depends on antenna construction and geometry. The peak response (`resonance') is typically obtained at an in-air frequency $f=c/L$, with $L$ the characteristic length scale of the antenna itself, and should occur when the antenna currents form standing waves along the length of the antenna, implying resonant frequencies in a vacuum $f_{res}=mc_0/2L$, with $m$ some integer. For a center-fed fat dipole antenna (such as those used by RNO-G), the resonant current amplitude must be maximal at the feed-point connecting the antenna to the cable conveying signal, corresponding to $m$ odd. The antenna used in this work has two such in-air resonances in the sensitive frequency band, at $\sim \qty{250}{\MHz}$ ($m=1$) and also $\sim\qty{750}{\MHz}$ ($m=3$).

The inductive, capacitive, and resistive characteristics of antennas determine their complex impedance $Z_A$ at a given frequency. The mismatch between the impedance of an antenna and the impedance of the input/output at the feed-point (where the antenna attaches to the cable [`transmission line'] conveying the signal in/out of the antenna to/from a measurement device such as a Vector Network Analyzer [VNA]) determines how well it will radiate/respond at a given frequency. The complex `reflection coefficient' $\Gamma$ is the ratio of the amplitude of the reflected wave and the incident wave in an antenna (similar to the general case of electromagnetic radiation being reflected/transmitted at a dielectric boundary) and provides a measure of the frequency-dependent antenna response; $\Gamma$ is defined as 
\begin{equation}
    \Gamma = \frac{Z_A - Z_0}{Z_A+Z_0},
\end{equation}
with $Z_0$ the transmission line impedance (usually $50\Omega$ for standard coaxial cable and purely real).
The magnitude of the complex 
$|S_{11}|$ parameter is related to the magnitude of the reflection coefficient via:
\begin{equation}
    |S_{11}|(dB) = 10\log_{10}(|\Gamma|),
\end{equation}
so $|S_{11}|$=0 dB corresponds to complete power reflection and therefore zero transmission; conversely, $|S_{11}|\to-\infty$ dB corresponds to perfect radiation at a given frequency.\footnote{For the remainder of this article, $S_{11}$ denotes the magnitude of the $|S_{11}|$ parameter and not the complex value.}

The medium into which an antenna is embedded can be characterized by its relative permittivity $\varepsilon_r$, relative permeability $\mu_r$, and conductivity $\sigma$.
The {\it in situ} media complex impedance $Z$ then varies as:
\begin{equation}
    Z = \sqrt{\frac{\mu}{\varepsilon}} = \sqrt{\frac{\mu_o \mu_r}{\varepsilon_0\varepsilon_r}} = Z_v \sqrt{\frac{1}{\varepsilon_r}} = \frac{Z_v}{n}
\end{equation}
where $\mu_0$ is the relative permeability of the vacuum, $\varepsilon_0$ is the relative permittivity of vacuum, and $Z_v$ is the free-space vacuum impedance (377$\Omega$). We assume that the media of interest to this work are non-magnetic ($\mu_r = 1$). Qualitatively, this permittivity dependence implies that an antenna with measured in-air frequency-dependent impedance $Z(\omega)$ evolves to an in-medium impedance as: $Z(\omega)\to (Z/n)(\omega/n)$, separately for both the real and imaginary components of $Z$.
In our case, for which a vertically-oriented antenna is deployed into a dry borehole, the antenna is not completely embedded in a single medium. Instead, it is immersed in a medium with an effective index of refraction $\nX{eff}$, which is a weighted combination of the $n\approx$1 air in the borehole and ${\it n}_{\text{firn}}$ of the surrounding ice, so ${\it n}_{\text{air}}\leq {\it n}_{\text{eff}}\leq {\it n}_{\text{firn}}$. Additional secondary effects like temperature, moisture content, and impurities are assumed to be negligible for this measurement. 

In this 
paper, we conduct experiments to demonstrate that measurements of this resonant 
frequency shift can be used to infer the local refractive index profile.

\section{Summit Station Measurements}

\subsection{Initial Measurement Campaign (2024)}
\begin{figure}[h]
    \centering   \includegraphics[width=\linewidth]{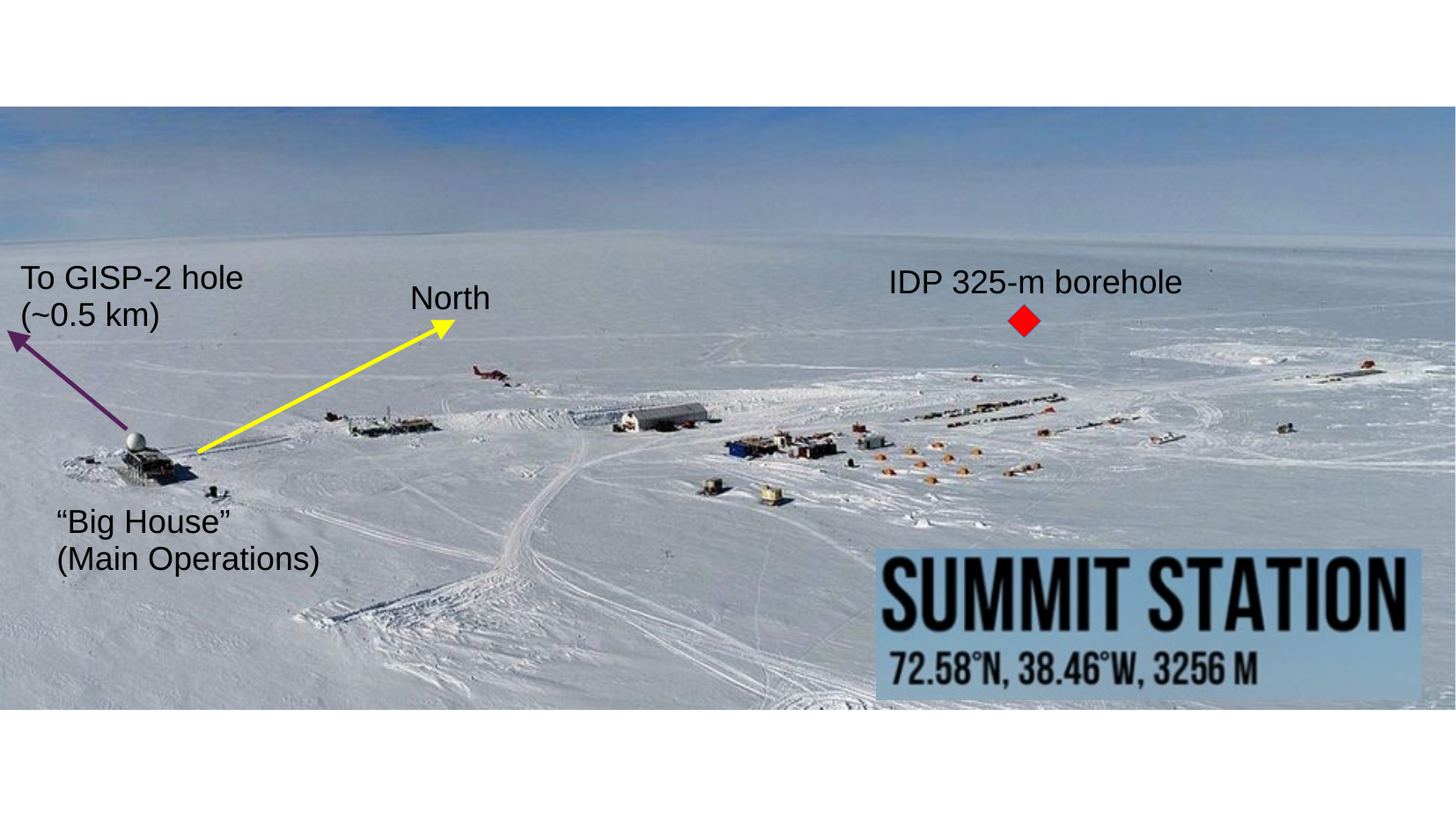}
    \caption{Location of 2024 IDP hole at Summit Station.}
    \label{fig:ssmap}
\end{figure}

\begin{figure}
    \centering
        \includegraphics[width=\linewidth]{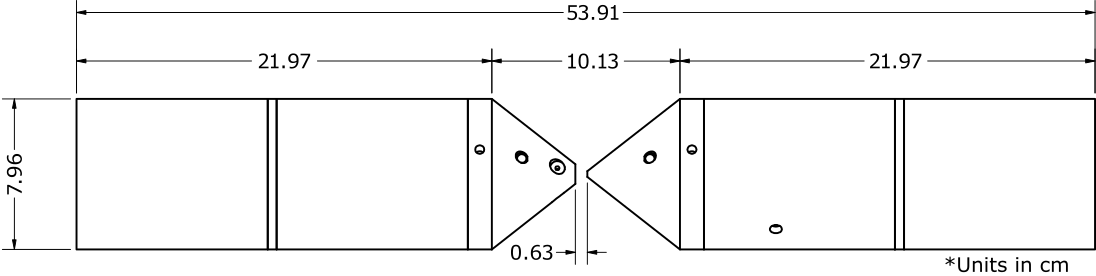} \\
    \includegraphics[width=\linewidth]{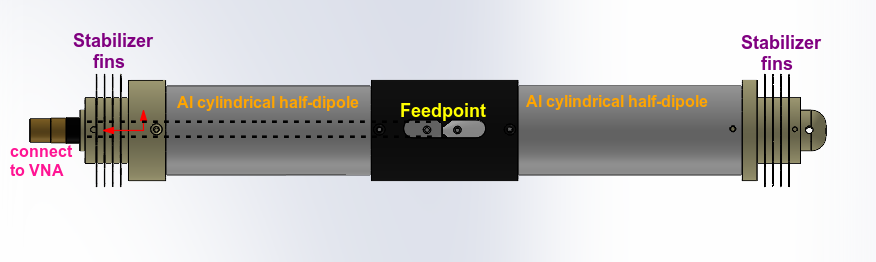}
    \caption{Top: KU-VPol antenna schematic showing dimensions of antenna employed for data-taking and feed-point detail. An N-connectorized cable is threaded through the left cylindrical chamber to the central feed-point and then connected to the Vector Network Analyzer. This antenna is similar in design to VPol antennas used for main RNO-G data-taking. Bottom: CAD model of same dipole antenna, illustrating central collar used to maintain desired spacing between cylindrical halves, and also showing endcap axial stabilizer fins, used for 2025 measurements.}
    \label{fig:cadmodel}
\end{figure}

\begin{figure*}[h]
    \centering
    \includegraphics[width=0.34\linewidth]{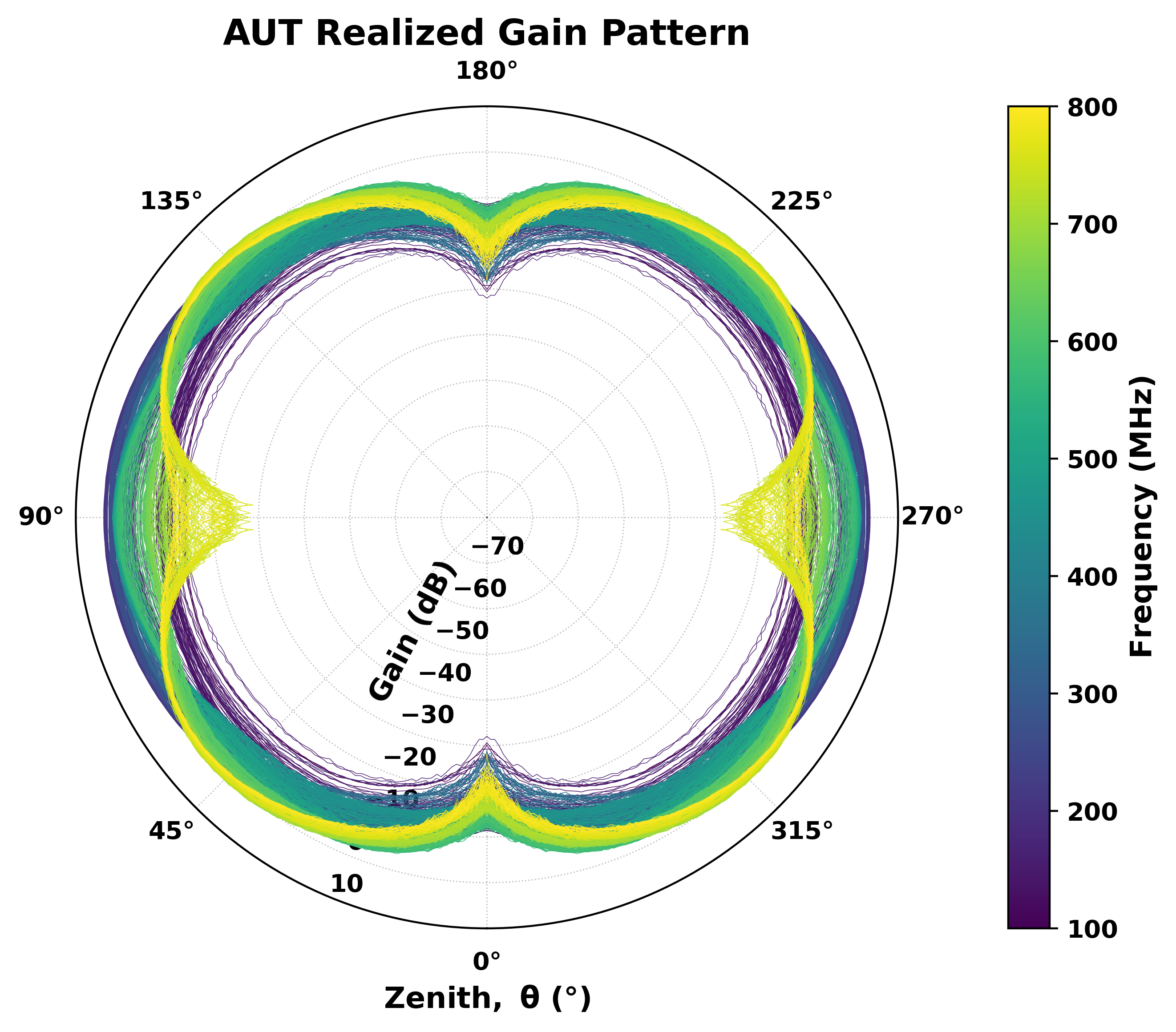}   \includegraphics[width=0.34\linewidth]{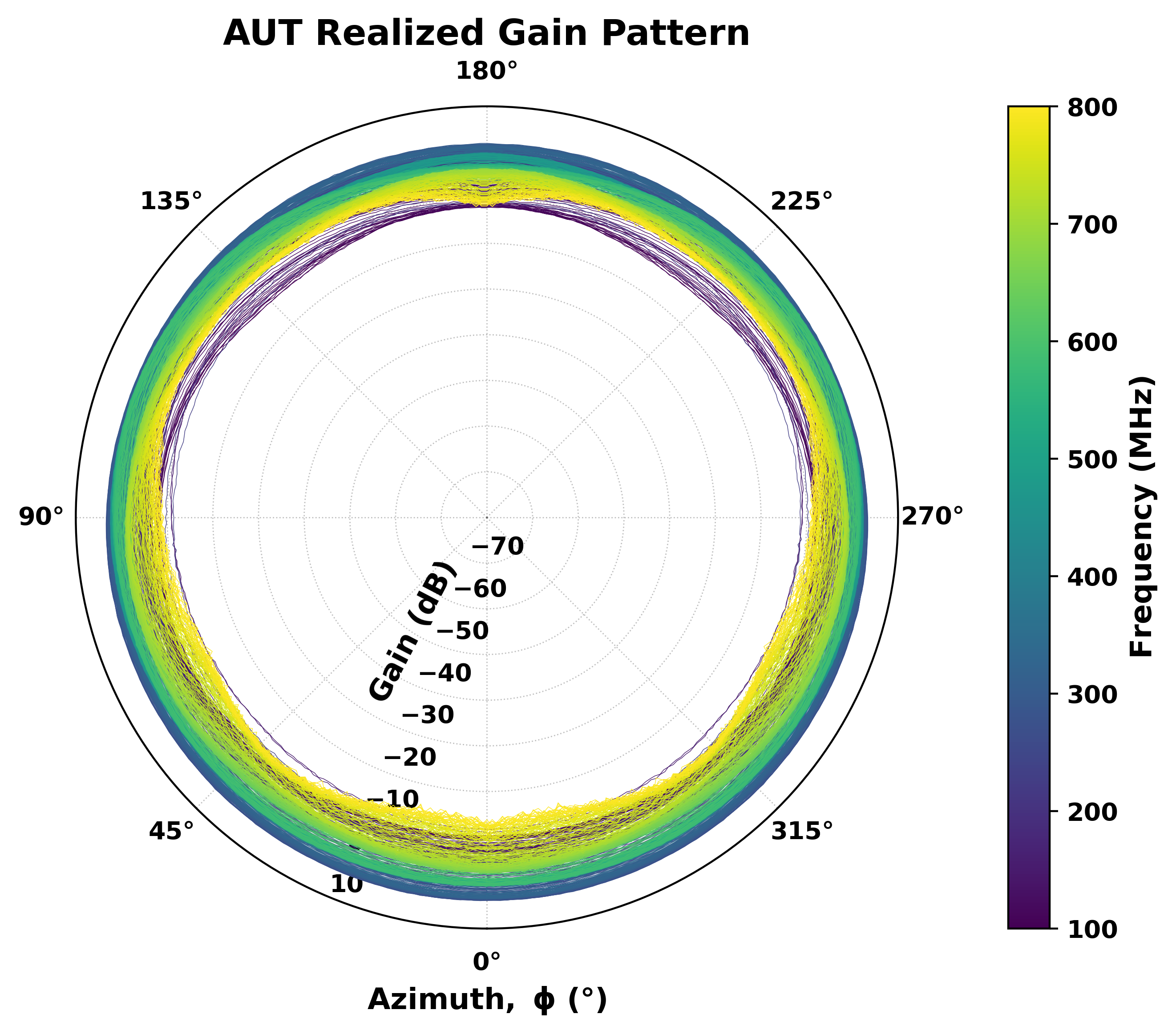}
    \caption{Beam pattern of dipole antenna used for data-taking, showing elevation (left) and azimuthal (right) gain. Measurements were taken in KU anechoic chamber.}
    \label{fig:KUDipoleGain}
\end{figure*}

A $\qty{350}{\m}$ deep, 98$\pm$1 mm diameter\footnote{We thank Jay Johnson of the Ice Drilling Program at UW, Madison for providing these details.} ice hole 
drilled at Summit Station by the Ice Drilling Program (IDP, based at UW, Madison) using the `700 drill' in June/July, 2024 for PI N. Saltzman. The location of the IDP-drilled hole, relative to other landmarks at Summit Station is indicated in Fig. \ref{fig:ssmap}.
This borehole provided an opportunity to collect antenna impedance data as a function of depth.
For measuring the antenna response to a depth of $\qty{100}{\m}$, we deployed a fat vertically polarized antenna (VPol; Fig. \ref{fig:cadmodel}) 
constructed at the University of Kansas (KU), based on the previous design used for transmitter measurements taken at the South Pole Ice Core Experiment (SPICE) borehole in December 2018 \citep{allison2020long}. Typical of dipole antennas, the in-air antenna beam pattern favors reception in the mid-plane (aka `boresight', and transverse to the long axis of the antenna) with a maximum gain of +3 dBi, falling off approximately as $\cos\theta$ with elevation angle $\theta$, and approximately uniform in azimuth ($\phi$), as illustrated in Fig. \ref{fig:KUDipoleGain}. 

As the antenna was lowered, reflection coefficient data were collected at $\qty{1}{\m}$ increments using a handheld FieldFox N9913B vector network analyzer (VNA).
For these measurements, the VNA frequency resolution was
set to 1 MHz. Although the highest possible frequency resolution (10000 points over 1 GHz frequency range, corresponding to 100 kHz resolution) was desirable, the limited battery lifetime of the N9913B FieldFox VNA imposed a practical limitation on frequency resolution, since data-taking time scales with the number of frequency bins. We note that the intrinsic frequency resolution of this instrument is quoted in the manual to be less than 1 Hz.\footnote{https://www.keysight.com/us/en/assets/7018-06516/data-sheets/5992-3702.pdf} After calibration (done prior to every set of measurements), the FieldFox N9913B VNA has a systematic frequency uncertainty of $\pm$0.5 ppm (parts per million), such that, e.g., at 1 GHz, the actual frequency uncertainty is 500 Hz. 
Within the limits of our resolution, we verified that the VNA reproduced the same $S_{11}$(frequency) profile in 16 separate S11 measurements, with no measurable scatter.

The manufacturer's claimed depth accuracy of the GV530 winch\footnote{geovista.co.uk} used to lower the antenna in 2024, for which the antenna signal cable was snaked around the winch cable, is $\qty{1}{\cm}$ (with no hysteresis), so error and uncertainty in the height of those measurements are considered to be negligible.

\begin{figure*}    \centering   \includegraphics[width=\linewidth]{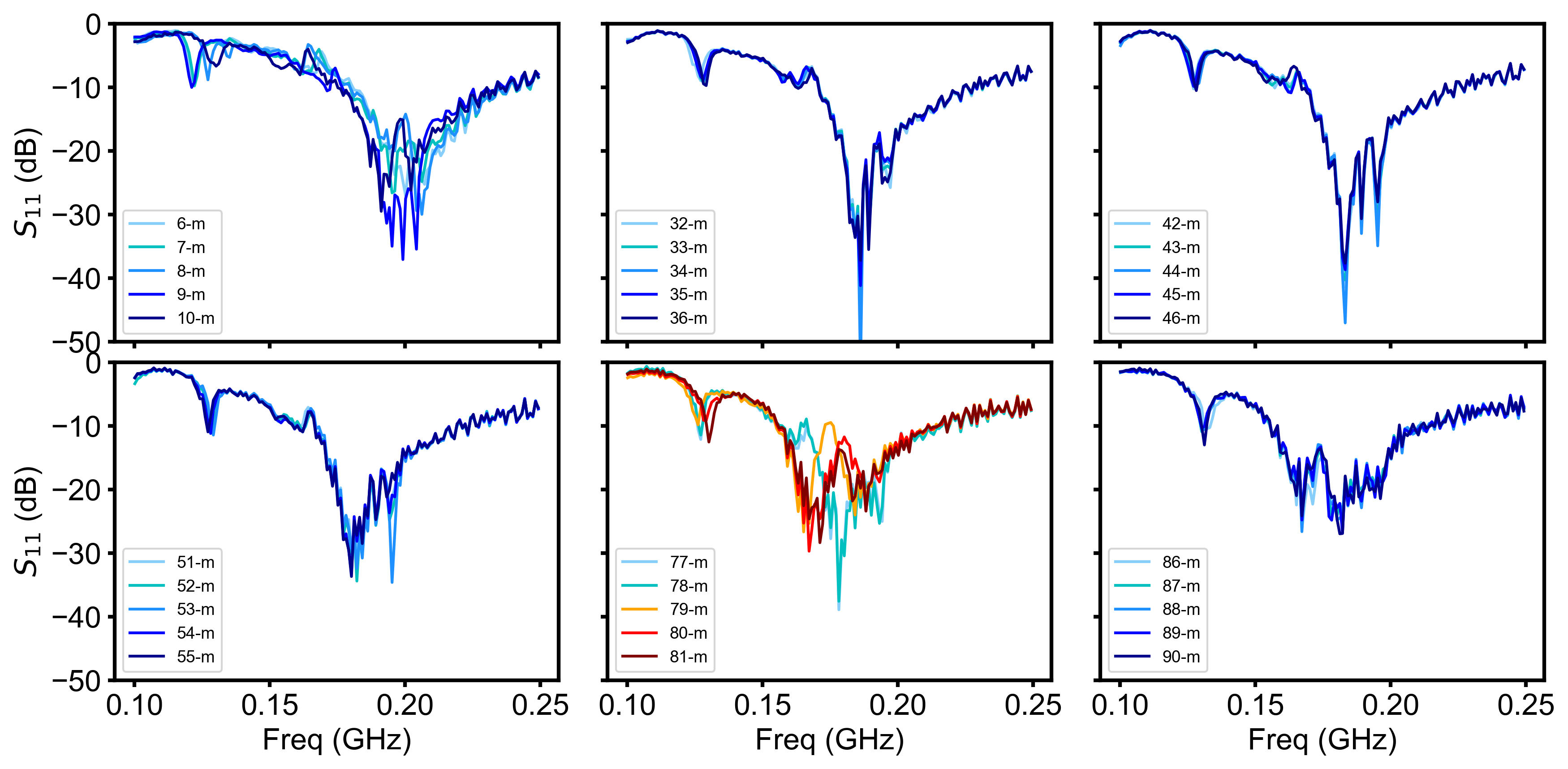}    \caption{Measured 2024 $S_{11}$ at various depths in ice, illustrating anomalous response observed at depths near 80 m (bottom row, central panel). Note the abrupt change in the shape of the resonance.} 
\label{fig:response_grid}\end{figure*}

Fig. \ref{fig:response_grid} displays some of the recorded 2024 $S_{11}$ profiles. 
We note an apparently anomalous response in the 2024 $S_{11}$ data appearing at $\qty{79}{\m}$ and then gradually decreasing over the next $\qty{5}{m}$ (discussed below).

Additionally, at depths shallower than $\qty{20}{\m}$ we observe significant scatter in the $S_{11}$ data points, indicating either some instrumental effect or true density fluctuations in this shallow depth interval.
We also observe coherent reflections from small impedance mismatches, becoming increasingly visible at high frequencies as a low-amplitude ripple in the $S_{11}$ data, and particularly prominent at high frequencies. The period of this ripple was observed to scale, as expected, on the length of cable used for the measurement.

\subsection{Improved (2025) Measurements}
The 2024 measurements identified one clear systematic uncertainty (possible tilt of antenna in the ice hole) and (at least) two unresolved science questions, namely the source of the scatter in the depth interval near 80 meters and also shallower than 20 meters. To investigate these questions further and also extract a refractive index profile from the higher-frequency m=3 resonance, a second measurement campaign was conducted at Summit Station in May, 2025.

\begin{figure*}[h]
    \centering
        \includegraphics[width=0.3\linewidth]{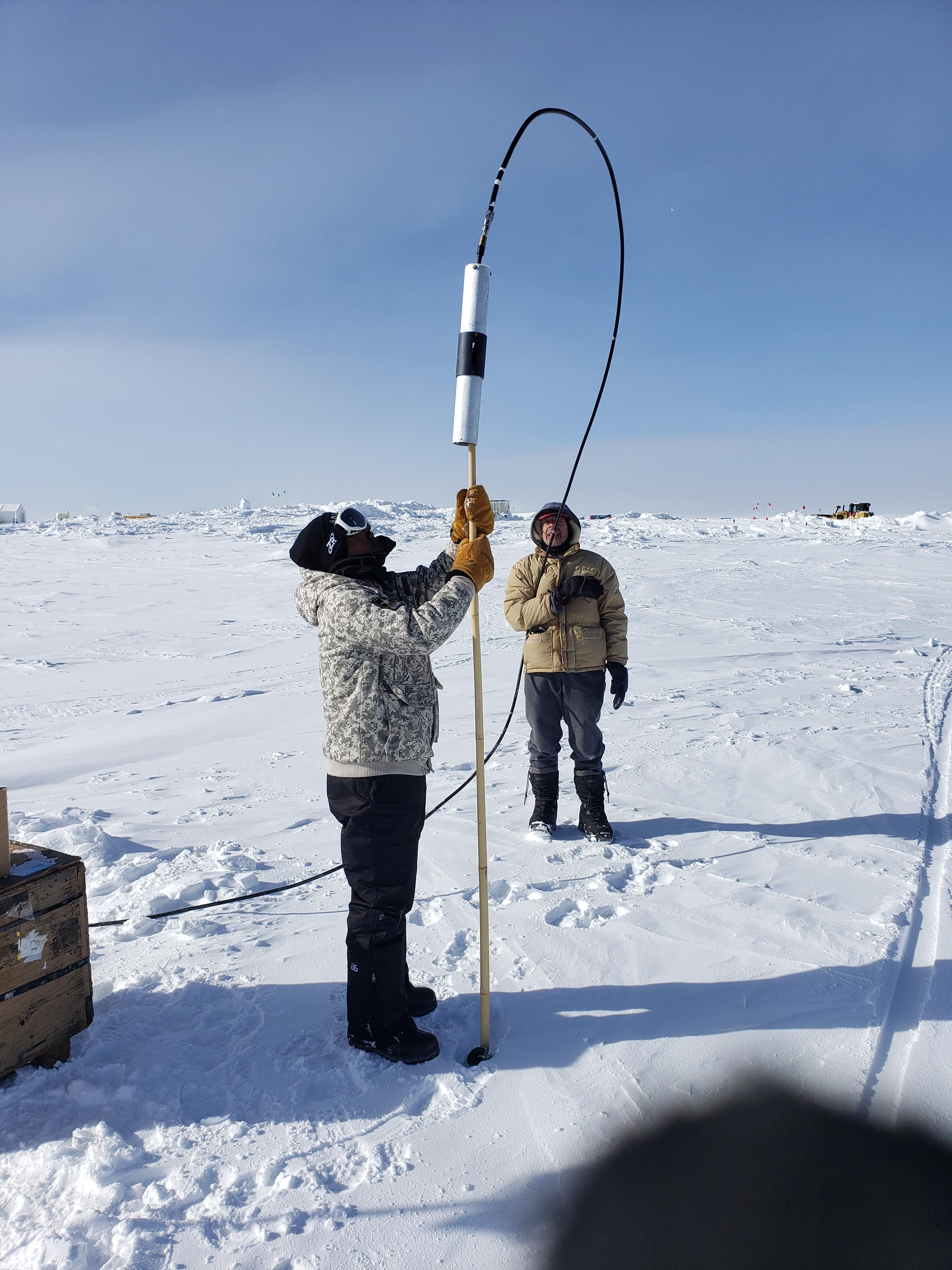}
        \includegraphics[width=0.3\linewidth]{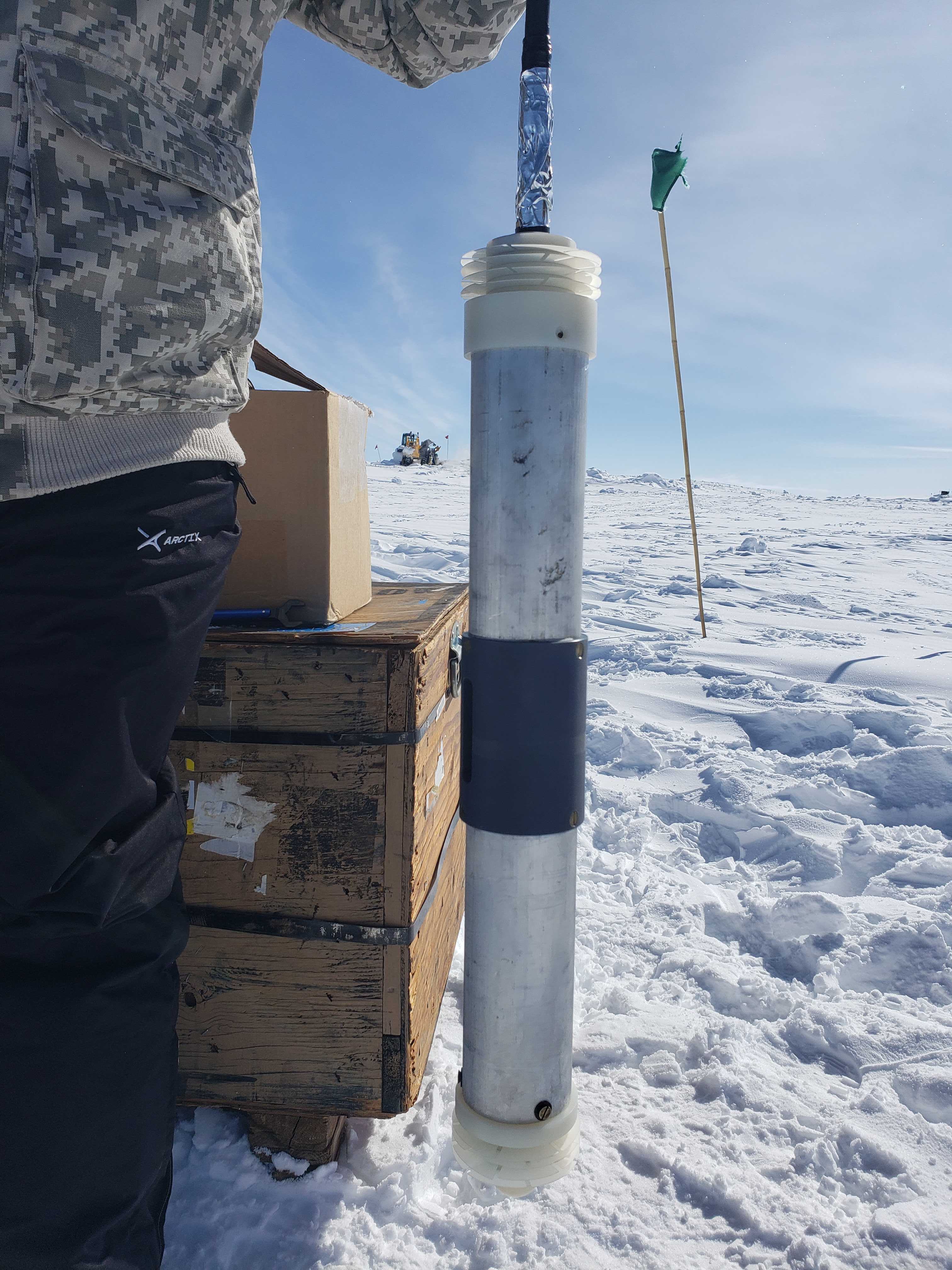}
    \includegraphics[width=0.32\linewidth]{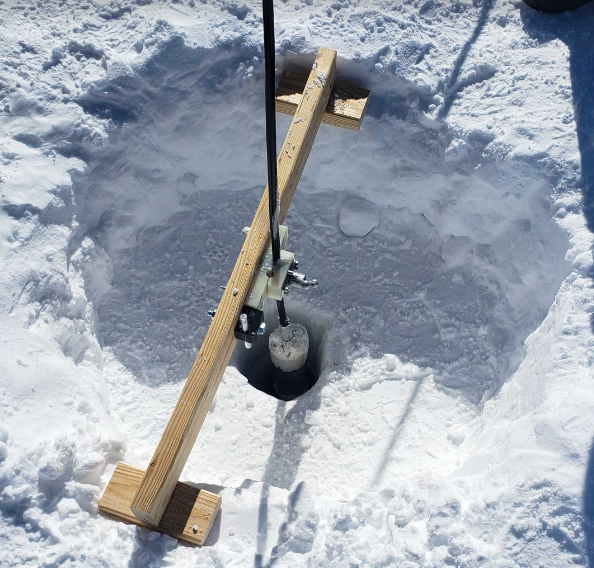}
    \caption{Pictures of dipole antenna (2025) at the IDP borehole. Top left shows in-air $S_{11}$ measurements, with Summit Station in background. Top right plot shows antenna at outset of data-taking; bottom illustrates large clearance of borehole at the surface, relative to antenna. Also shown is wooden brace used (2025 only) to ensure vertical orientation of antenna and also for strain relief (assumed to have no effect on measured $S_{11}$ after calibration). For 2024 measurements, signal cable was co-lowered with strength member cable from GeoVista GV530 winch (not shown), which was mounted atop a stationary wooden platform for stability.}
    \label{fig:2025pix}
\end{figure*}

Fig. \ref{fig:2025pix} shows the antenna at the outset of the 2025 measurements; evident in the picture is the large clearance of the borehole (approximately twice nominal diameter) at the throat (unfortunately, a mechanical caliper was not available for mapping the throat profile).
In 2025 the antenna was lowered by hand to eliminate any possible RF emissions due to a proximal generator or winch, using lower-loss coaxial cable (to improve the Signal-to-Noise Ratio [SNR] at higher frequencies), to a depth of $\qty{85.5}{\m}$ with an estimated uncertainty of $\qty{10}{\cm}$. Spacers introduced in 2025 to the endcaps of the antenna attempted to address the dominant uncertainty in the 2024 data analysis arising from unknown axial alignment (discussed in more detail below). 
VNA sweep averaging ($N_{avg}$=16 sweeps per data record) in 2025 also improved the visibility of the higher-frequency $m$=3 ($f_2$) harmonic, in addition to the $m$=1 $\frac{c_0}{2L}$ ($f_1$) fundamental. For the 2025 dataset, measurements were taken as the antenna was both raised and lowered, providing cross-checking of the two data sets.
Data-taking parameters for the two years are compared in Table \ref{tab:data_taking}. 

\begin{table}[h]
    \centering
    \begin{tabular}{c|c|c}
 Parameter & 2024 & 2025 \\ \hline
 $N_{avg}$ & 1 & 16 \\
 coaxial cable & 120m LMR-400 & 30m LMR-600 \\
    &   &+2$\times$30m LDF4-50A \\
    antenna lowering & GV530 winch & manual \\
    depth uncertainty $\sigma_z$ & 1 cm & 10 cm \\
 VNA output power & -15 dBm & 0 dBm \\
 Outside Air Temp & -8 C & -25 C \\
 Maximum depth & 100 m & 85.5 m \\
 Axial alignment & none & endcap spacers \\
 Data Taking & $\downarrow$ only & $\downarrow$ and $\uparrow$ \\
 Resonances Fitted & $f_1$  & $f_1$ and $f_2$ \\
Parameterization & Double Gaussian & Double Lorentzian \\
\end{tabular}
    \caption{Data-taking parameters for the 2024 vs. 2025 datasets.}
    \label{tab:data_taking}
\end{table}

\subsection{Extraction of $S_{11}$ resonant frequencies from raw $S_{11}$ data}

We extract, from data, the resonant frequency of the antenna in the IDP hole as a function of depth $f_r(z)$, by fitting the $S_{11}$ distribution at each depth to either a double-Gaussian (2024 data), defined as 
\begin{equation}
        P(f) = A_1\exp\bigg(\frac{-(f-\eta_1)^2}{2\sigma_1^2}\bigg)+A_2\exp\bigg(\frac{-(f-\eta_2)^2}{2\sigma_2^2}\bigg)
\end{equation}

with $A_{1,2}$ the amplitude of the Gaussian, $\eta_{1,2}$ the mean and $\sigma_{1,2}$ the standard deviation, or a 
double-Lorentzian (2025 data):
\begin{equation}
        P(f) = \frac{A_1}{(f-\eta_1)^2+\gamma_1^2}+\frac{A_2}{(f-\eta_2)^2+\gamma_2^2}
\end{equation}
with $\gamma_{1,2}$ is the width. 

\begin{figure}
\centering
    \includegraphics[width=\linewidth]{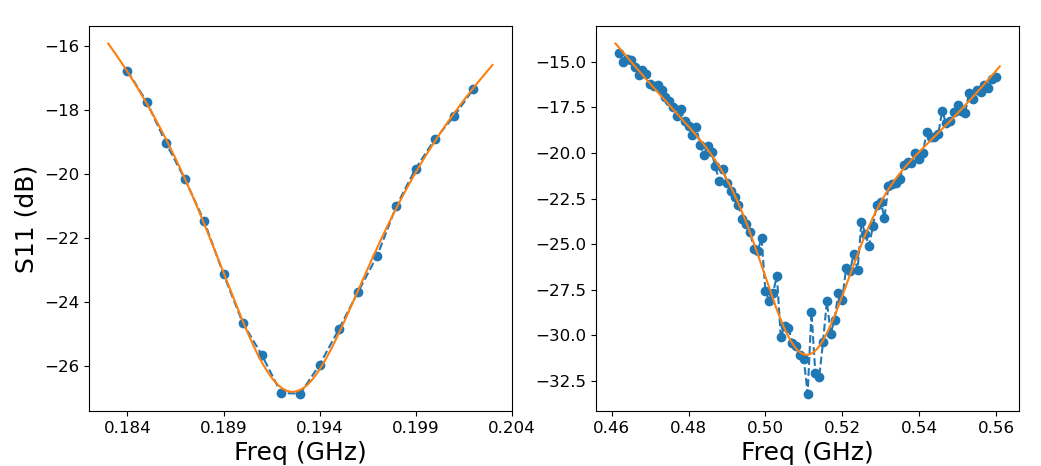}
    \caption{Double Lorentzian fit to 2025 $f_1$ $S_{11}$ profile (left) and double Lorentzian fit to 2025 $f_2$ $S_{11}$ distribution using data taken at a depth of $\qty{13}{\m}$ (selected arbitrarily). }
    \label{fig:Example_fit}
\end{figure}

\begin{figure}[h]
    \centering
    \includegraphics[width=\linewidth]{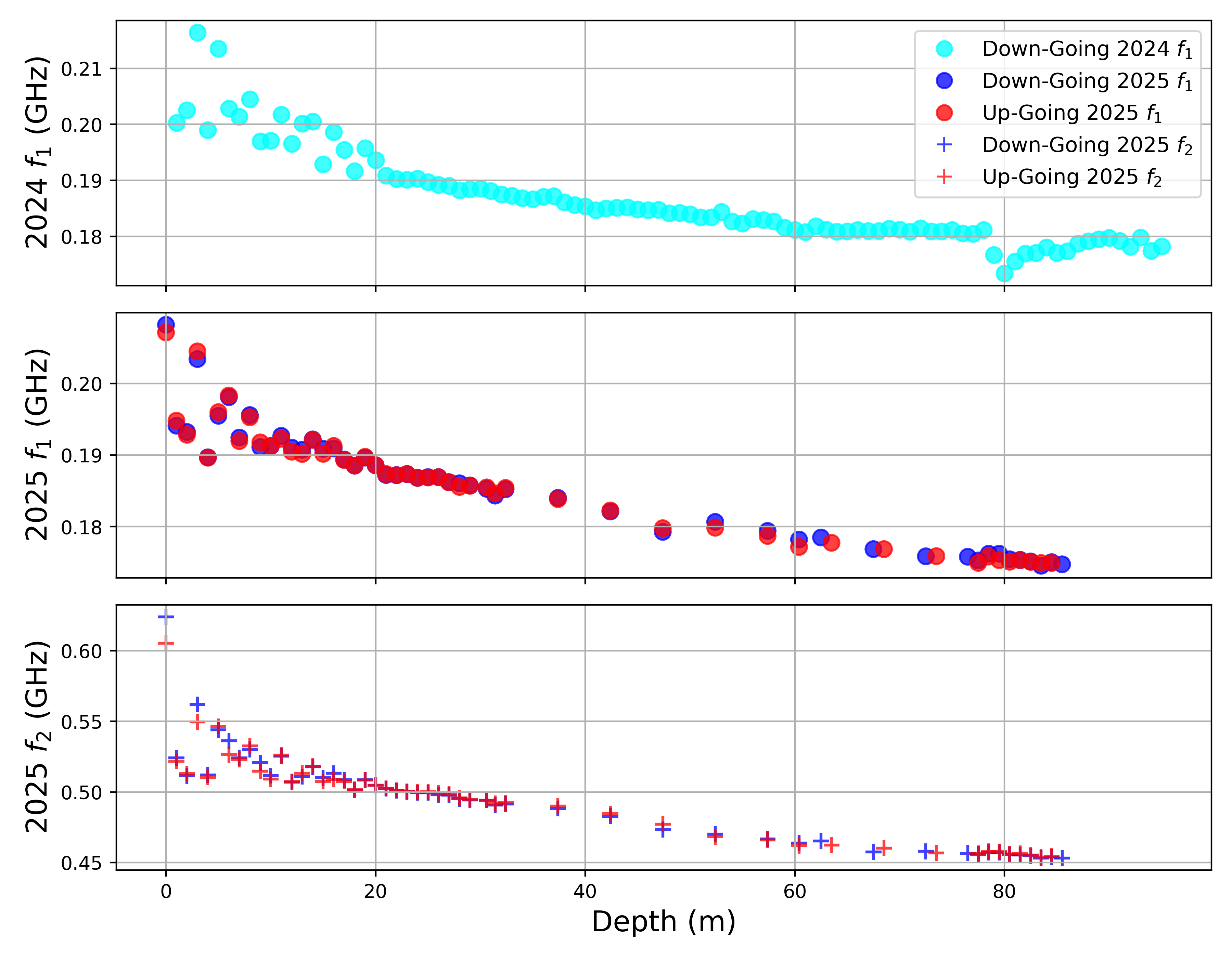}
    \caption{Raw extracted resonant frequencies, as returned from double-Gaussian/double-Lorentizian fit, for each indicated data set, plotted against antenna depth.}
    \label{fig:data-resonance}
\end{figure}

For m=1 (2024 data), the double Gaussian was fit to $S_{11}$ over the range $\qty{1}{\MHz}$ to $\qty{300}{MHz}$. Sample fits are shown in Fig. \ref{fig:Example_fit}.
Additional $f_{res}$ estimators were examined (tracking the $S_{11}$ minimum value, calculating the minimum of a weighted sum, or using an asymmetric Laplacian form, e.g.) and found to yield similar results, though with less fitting stability and/or poorer convergence characteristics compared to the double Gaussian/Lorentzian. The extracted resonance values $f_{res}(z)$ for our data sets are shown in Fig. \ref{fig:data-resonance}.

\begin{figure}[h]
    \centering
    \includegraphics[width=\linewidth]{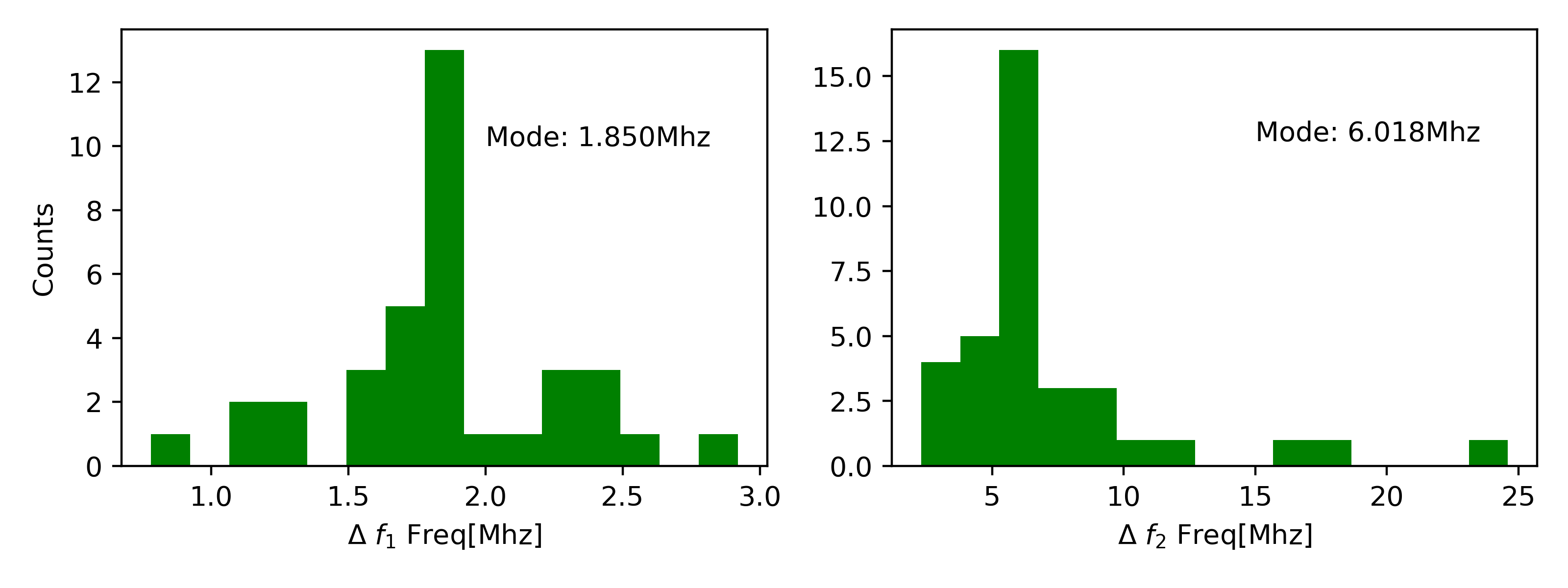}
    \caption{Difference of $(f_{\uparrow})$ and $(f_{\downarrow})$ for $f_1$ and $f_2$.  respectively (2025 data-taking). The mode bin is labeled.}
    \label{fig:diff_residual}
\end{figure}

In the 2025 data there is a clear 2-3 MHz systematic reduction of the resonant values obtained during the (later) upwards-going (designated $f_{\uparrow}$) compared to the previous \\downwards-going data $(f_{\downarrow})$ for both $f_1$ and $f_2$, particularly for $z<\qty{50}{m}$. Several possible causes for this discrepancy were considered. In the field, accumulated snow was observed both above and also between each of of the endcap stabilizing fins when the antenna was raised from the borehole after data-taking, presumably due to the antenna scraping the side of the borehole while being lowered/raised. Such snow could affect the measured resonance by raising the effective refractive index around the antenna (by displacing air with snow), resulting in a lowered measured resonant frequency. Additionally, there could be VNA temperature dependence (expected to become more obvious with time), as the 2025 data were taken while the ambient air temperature was --25 C, with a wind chill of --60 C. Both of these could bias later VNA readings towards lower values (although we were unable to intentionally reproduce a systematic, temperature-dependent droop of the VNA response in the lab). We empirically account for this by taking the mode of a distribution of $(f_\uparrow(z)-f_\downarrow(z))$ (Fig. \ref{fig:diff_residual}) and applying that offset to the $f_\uparrow$ data.

\section{Analysis}
\subsection{Measurement Strategy}
From our reflection coefficient measurements, we calculate the variation of the m=1 fundamental resonant frequency (denoted $f_1$) with depth for data taken in both 2024 and 2025, as well as the m=3 second resonance (first harmonic, and denoted $f_2$) for 2025, to derive the desired refractive index profile.
We must therefore translate our $S_{11}(z)$ measurements (made in a medium with local effective refractive index ${\it n}_{\text{eff}}(z)$) into an ice \RIP profile; this requires that our frequency dependence be re-formulated as a refractive index dependence, i.e., $S_{11}(z)\to S_{11}(n(z))$, which we can then invert to obtain the desired $n(z)$. We expect the fundamental resonance (m=1) of the antenna to scale as $f_{m=1} \sim c_0/(2\lambda_{m=1})$, where $\lambda_{m=1}$ is the m=1 fundamental wavelength. Since $\lambda_1$ is set by our (fixed) antenna length, and since $c$ scales as 1/$n$, the resonant frequency should also scale as 1/$n$. We correspondingly assume the functional form $f(n)=a/(b+n)$, with the constant $a$ primarily determined by the antenna length, and the constant $b$ allowing for the possibility of an offset independent of $n$.

\subsection{Determination of RIP from frequency data}
\begin{figure*}[h]
    \centering
    \includegraphics[width=\linewidth]{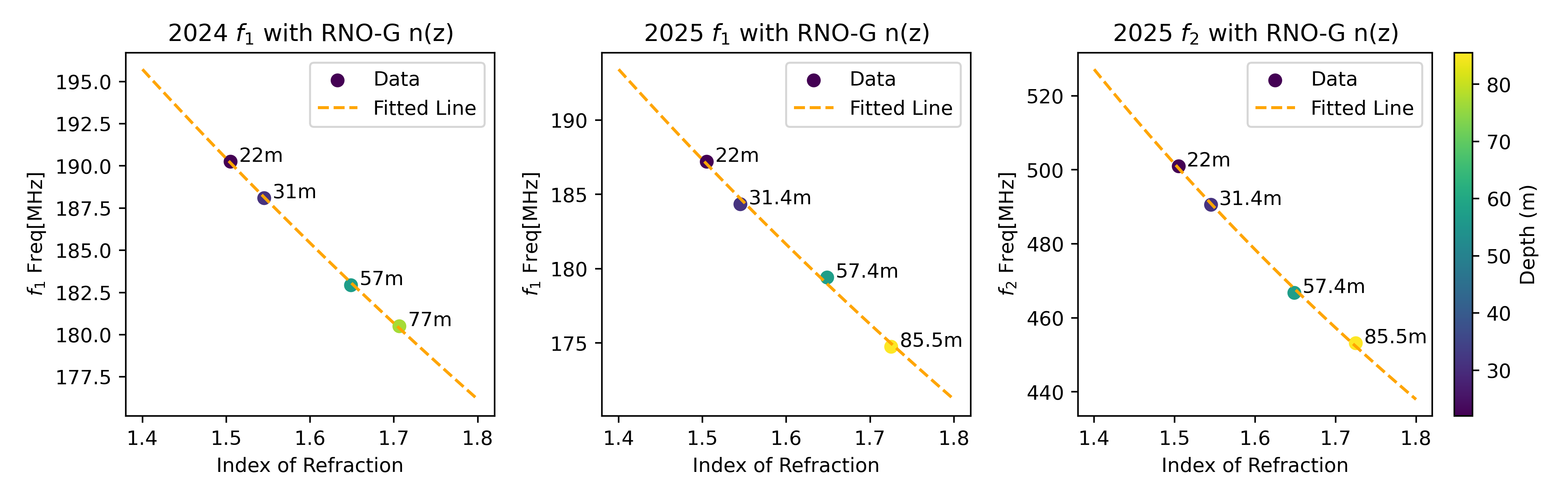}
    \caption{Fits to $f(n)$ for 2024 $f_1$, 2025 $f_1$, and 2025 $f_2$ data sets (detailed below), using four depth points selected from the `smooth' depth region. }
    \label{fig:nvf}
\end{figure*}

 To extract the constants $a$ and $b$ in our parameterization for $f_{res}(n)$, we plot the measured resonant frequencies against the refractive index implied by the independent density data, described in \cite{RNOG_Philipp}.
Four points (two at the endpoints of the interval, and two intermediate points) are selected over the depth interval (22 m -- 80 m) for which the frequency data are `smooth'; we find that our $f(n)=a/(b+n)$ ansatz gives a reasonable match to the observed NPM density data (Fig. \ref{fig:nvf}). 
 Having calibrated our \RIP functional form, we can then extrapolate into the 
 shallow firn and quantify refractive index fluctuations, independent of the calibration data set; we later check against gravimetric measurements made on cores taken from the same hole.

Table \ref{tab:ab} summarizes the values of $a$ and $b$ we extract from our fits to $f(n)=a/(b+n)$. To assess the goodness-of-fit of our $f(n)\sim 1/n$ ansatz, the second column from the right tabulates 
the average residual of our fit (`$<\delta>$'), relative to our data points (i.e., $\sqrt{\Sigma_i(f_{measured,i}-f_{fit})^2}$/N), where N is the number of points used in the fit, $f_{measured,i}$ is the fit resonant frequency for each depth value, and $f_{fit}$ is the value of frequency, for a given depth value returned by the fit. The right-most column shows, for comparison, the deviation obtained using
a linear fit $f(n)=a+bn$, which yields values typically larger by $\sim$20\%, indicating  preference for the inverse, rather than linear, scaling.
\begin{table*}[h]
\centering
\begin{tabular}{c|c|c|c|c|c}
Data Set & a [MHz] & b & $N_{pts}$ (all depths/22 m$\to$80 m) & $<\delta>$  [MHz] &  a+bn  $<\delta>$ [MHz]\\ \hline
2024 (down-only) & 742.27 & 2.39 & 95/73 & 0.030/0.018 & 0.036/0.017\\
2025, $f_1$ (up+down) & 596.46 & 1.68 & 102/56  & 0.019/0.0086 & 0.023/0.0090\\
2025, $f_2$ (up+down) & 1035.45 & 0.56 & 102/56  & 0.021/0.0086 & 0.026/0.0092 \\ \hline
\end{tabular}
\caption{Fitted values of (a,b) and goodness-of-fit metrics. Last column verifies that a linear, rather than inverse dependence of frequency with refractive index results in a poorer fit.}
\label{tab:ab}
\end{table*}

To estimate our systematic error due to our selection of the four fit points, we also compared with a fit using all the data points and tabulated the difference in the extracted RIP values. Deviations are typically 0.01 units of $n$.

\begin{figure*}[h]
    \centering
    \includegraphics[width=0.85\linewidth]{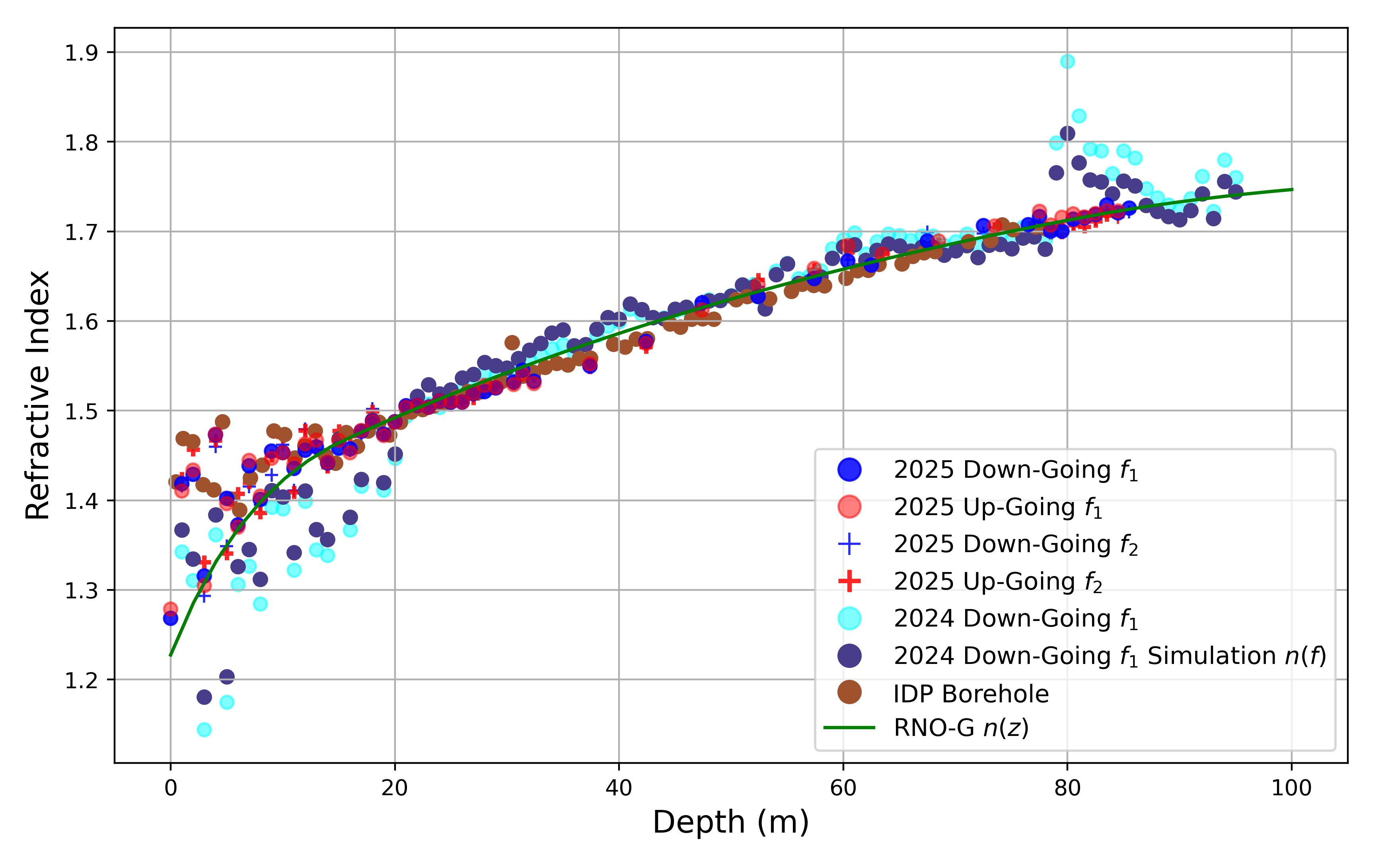}
    \caption{Overlaid $n(z)$ profiles for each dataset as well as preferred RNO-G n(z) parameterization from \citep{RNOG_Philipp}. `IDP Borehole' data has slightly slightly different binning refering to density data taken from the IDP hole, converted to refractive index assuming the linear scaling $n(z)=1+0.8485\rho(z)$.
    }
    \label{fig:n(z)-profile}
\end{figure*}

Using the extracted values of $a$ and $b$ for each data set, we can now translate our tabulated values of resonant frequency as a function of depth $f_{res}(z)$ from Summit data 
to obtain the $n(z)$ profiles shown in Fig. \ref{fig:n(z)-profile}. 
Recently, gravimetric density data for the IDP borehole were made available to our group; this provides a powerful cross-check of our $n(z)$ extraction procedure.\footnote{We are indebted to Murat Aydin, Eric Saltzman and John Patterson for supplying these measurements.} Those data, in general, show good agreement with our 2025 measurements, as shown in Fig. \ref{fig:n(z)-profile}.

\begin{figure}[tp] 
\centering
        \includegraphics[width=\linewidth]{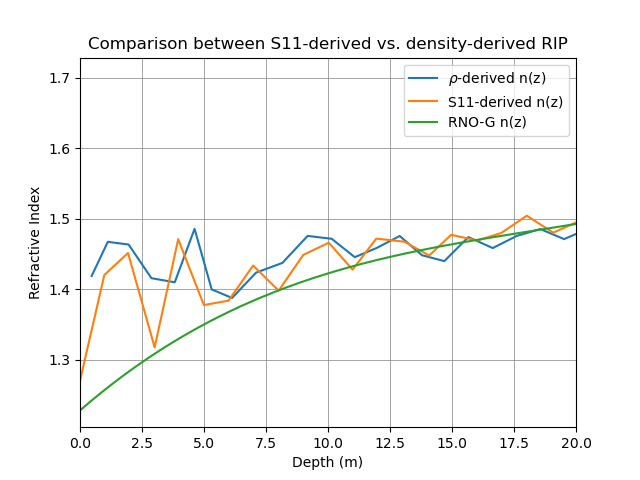}
    \includegraphics[width=0.495\linewidth]{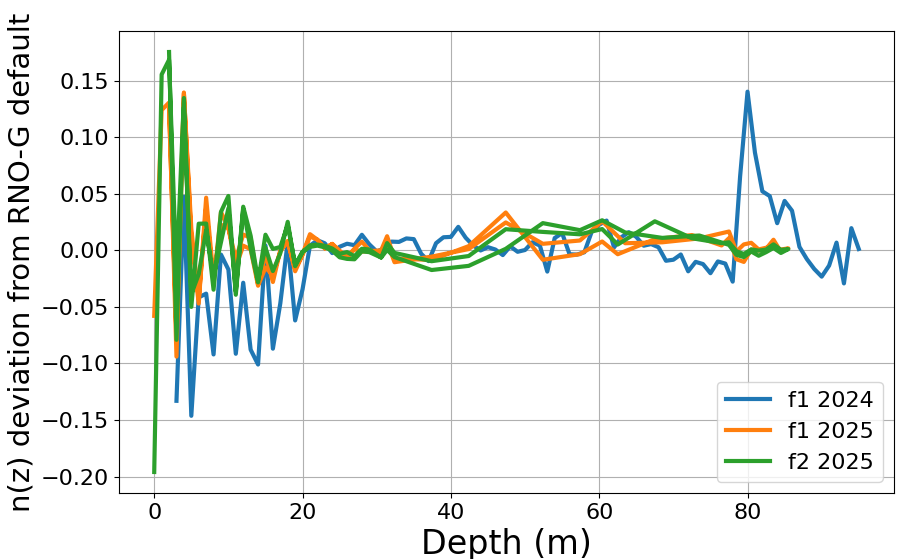}
    \includegraphics[width=0.495\linewidth]{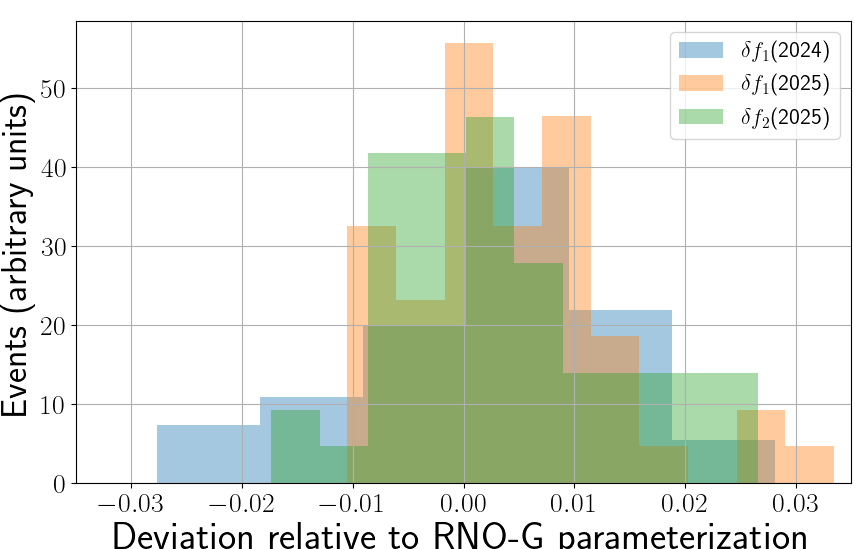}
    \caption{Top: Comparison of $n(z)$ profile based on IDP-borehole density data with $n(z)$ profile based on $S_{11}$ IDP-borehole measurements, for shallow depths. Bottom Left: Deviation between indicated $n(z)$ profile, relative to RNO-G default parameterization.  Bottom Right: One dimensional projection of residuals in previous plot over interval from 20--80 m; means and widths of distributions are: $<\delta f_1(2024)>=(3.5\pm1.4)\times 10^{-3}$, 
    $<\delta f_1(2024)>=(4.1\pm3.6)\times 10^{-3}$, and
    $<\delta f_2(2025)>=(3.2\pm4.4)\times 10^{-3}$. We interpret the general consistency with zero offset as lending support to the assumption of inverse RIP scaling with frequency, used to extract the $n(z)$ profile.}
    \label{fig:nzdev}
\end{figure}

Restricted to the shallow depth region ($z<$20 m), we observe that the density-derived IDP hole RIP visually tracks our $S_{11}$-derived RIP (Fig. \ref{fig:nzdev}, top), averaging $f_{1,\uparrow}$, 
$f_{1,\downarrow}$, $f_{2,\uparrow}$, and 
$f_{2,\downarrow}$), indicating that both techniques are sensitive to real density fluctuations in the shallow firn.

We have performed additional cross-checks of our
functional parameterization for $f(n)$. The bottom left panel of Figure \ref{fig:nzdev} shows the deviation between our extracted $n(z)$ values relative to the RNO-G parameterization; the bottom right panel displays the distribution of deviations for the three data sets considered. Overall, in the `good' region (20--80 m), we observe that the deviations, relative to the $n(z)$ parameterization are flat (and therefore unbiased in depth); the magnitude of the mean of the deviations is of order 0.003 units of $n$.
 
\begin{figure}[tp]
    \centering
    \includegraphics[width=\linewidth]{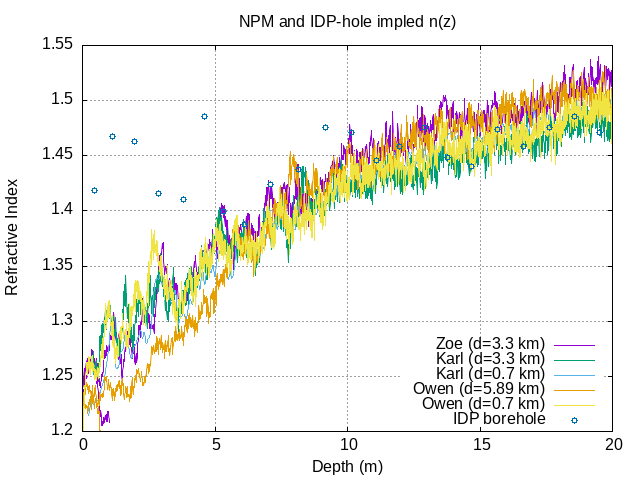}
    \caption{Comparison of $n(z)$ profiles implied by Hawley\& Morris (private communication) and \citep{morris2003density,Hawley_Morris_2006} $\rho(z)$ NPM data taken in the vicinity of Summit Station (retaining original site-naming convention; these data are one input to the $n(z)$ parameterization favored by RNO-G). Also included are density data taken at IDP hole in 2024. Distances indicated in key are relative to Summit Station main base. At shallow depths, we note significant scatter between sites separated by less than 1 km laterally.}
    \label{fig:NPMnvz}
\end{figure}

For comparison, we have tabulated the neutron probe monitor density data taken in the vicinity of Summit, and translated to refractive index for comparison (Fig. \ref{fig:NPMnvz}).\footnote{We are indebted to Liz Morris for making these data available for our use.} The 2024 IDP density data are also overlaid. The profile shapes generally track each other; we note significant variations in the calculated refractive index profile over sub-km spatial scales, comparable to the separation between the 350-m IDP hole and the nominal Summit Station NPM data-taking site, underscoring the importance of local measurements. Similar to the trend observed in our 2025 $n(z)$ profile, the IDP borehole density data overshoots the NPM data in the shallow depth regime. 

\subsection{Consistency between 2024 and 2025 data sets and Uncertainties}
We have attempted to quantify the level of agreement between the August, 2024 and May, 2025 data samples. Although they are consistent in the intermediate depth interval (20--80 m), they conspicuously deviate at depths of 80--85 m, for which the 2024 $f_1$ data 'overshoot' the density-derived refractive index, and in the shallow firn (depths less than 22 meters). Whereas the 2024 data-taking employed a (large metal) winch at the top of the hole, with signal and winch cable both in the hole, the 2025 data-taking used a single coaxial cable, eschewing use of the winch and/or ancillary cables.  
Perhaps most importantly, there was no attempt to axially center the antenna in the initial 2024 data-taking. 

The comparison is also complicated by the 2--3 MHz systematic offsets between different data samples, as well as the possible effects of additional one year of snow accumulation at the IDP hole site (averaging 24 cm water-equivalent yearly, at Summit Station) and/or densification effects resulting from human activity, including heavy machinery, at the drill site in the intervening time period). Ignoring any such time-dependent effects, a direct Pearson's coefficient calculation gives a correlation coefficient of 0.537. Alternatively, 
we can allow for a depth offset (but less than one meter) between the two data samples, by calculating the binary coherence between the shapes of the two $n(z)$ curves, as measured by whether the $(i+1)^{th}$ data point is `coherently' higher/lower than the $i^{th}$ data point. We find 17 cases for which a point-to-point increase/decrease in the 2024 data coincides with a correlated increase/decrease in the 2025 data, compared to three instances where they are anti-correlated, giving a binary probability in excess of $>$99\% correlation in the shallow region. 
Importantly, we find that the $S_{11}$ fluctuations observed in the shallow depth interval are reproduced for the 2025 up-going vs. down-going data sets, indicating that these are real density variations and not instrumental effects.

\begin{figure*}
    \centering
    \includegraphics[width=.495\linewidth]{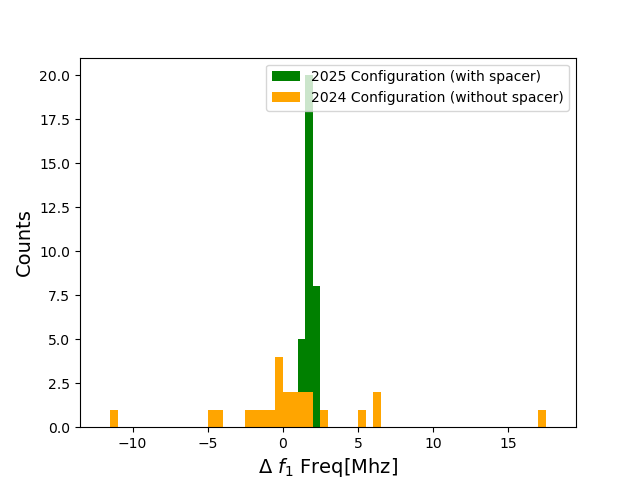}
        \includegraphics[width=.495\linewidth]{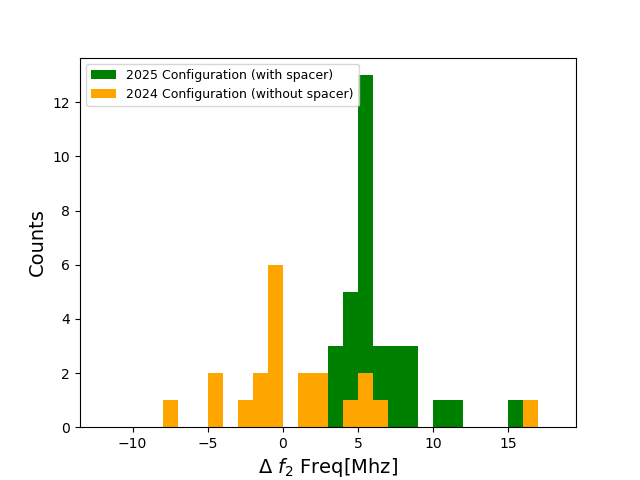}
    \caption{Comparison between $f_{1,\uparrow}-f_{1,\downarrow}$ frequencies extracted for runs (depths shallower than 20 m) using antenna spacers for centering/axial stability (2025 data configuration; green) vs. runs without stabilizers (2024 data configuration; orange). We observe increased consistency between the up/down data samples with spacers in place.}
    \label{fig:UpDownSpacers}
\end{figure*}

To investigate the effect of the 2025 stabilizer, we took dedicated runs in May, 2025, with and without the stabilizer, down to 20 meters, observing a narrower distribution with the stabilizer (Fig. \ref{fig:UpDownSpacers}) in place. We note that the diameter of boreholes are typically largest near the surface -- we measured a throat diameter of 21.8 cm of the borehole at the surface in 2025, indicating that off-axis effects, particularly near the surface, could be severe. Moreover, simulations indicate that, if the antenna were inclined on the hole wall, we should observe a systematic shift in resonant frequency (and therefore RI).

For 2024, each depth only had one measurement, rendering it unfeasible to estimate a true statistical uncertainty. Instead, we estimate the systematic error from the measuring device by scaling the uniform errors of the $S_{11}$ profile until the fit has a reduced chi-squared $\chi^2/\text{ndof} = 1$.\footnote{A fuller description of the chi-squared rescaling (standard in particle physics) can be found in the literature  \citep[Introduction (pp. 1--19)]{Cowan_2019, PhysRevD.98.030001}.} The Hessian matrix error on the returned fit $\eta_1$ value at a given depth was used as the VNA systematic uncertainty. This is only an approximate uncertainty as the double Gaussian parameterization is empirical, and not a first-principles description of the underlying $S_{11}$ trace distribution. Therefore, the uncertainties extracted from the fits should only be viewed as a proxy for the true statistical uncertainties of taking multiple measurements at the same depth. The uncertainty in the parameterization used to determine the depth-dependent resonant frequency can be assessed by considering the variation between the extracted $n(z)$ profile using three fitting parameterizations: i) the double Gaussian or ii) Lorentzian described earlier (our default), and iii) the Doniach-Sunjic (DS) line shape \citep{SDoniach_1970}, which is found to empirically give a reasonable match to our $S_{11}(f)$ line shapes. The deviation between the three fitting schemes has a center inter-quartile range of $\Delta n=0.0064$, indicating reasonable agreement. 
For the 2025 data set, the magnitude of errors is estimated directly from the discrepancy between up/down, at each data point.

There is a systematic uncertainty from the measurement device itself due to limited instrumental resolution. Other systematic errors inherent in the network analyzer include possible calibration drift introduced by any bends in the cable and or any temperature variation during data-taking (the VNA is nominally rated for operation between -10 C and +55 C).

We tabulate considered systematic uncertainties in Table \ref{tab:systematics}. 
\begin{table*}[h]
    \centering
    \begin{tabular}{l|ccc}
    \hline
        Source of Uncertainty& $\sigma_n$ ($\sigma_{f1}$) (2024) & $\sigma_{f1}$ (2025) & $\sigma_{f2}$ (2025) \\ \hline
        Alignment Uncertainty (Maximum) & 0.05 ($\qty{4}{\MHz}$) & 0.014 ($\qty{1.2}{\MHz}$) &  0.021 ($\qty{2.5}{\MHz}$) \\
        VNA Uncertainty & 0.012 ($\sim\qty{1}{\MHz}$) & 0.012 ($\sim\qty{1}{\MHz}$) & 0.012 ($\sim\qty{1}{\MHz}$)\\
        $S_{11}$ Fitting Scheme & 0.004 ($\qty{.35}{\MHz}$) & 0.004 ($\qty{.35}{\MHz}$)  & 0.007 ($\qty{0.6}{\MHz}$)  \\
        Uncertainty in (a,b) & 0.010 & 0.008 & 0.011 \\
        Uncertainty in $\rho$/RI relation & 0.003 & 0.003 & 0.003 \\
        Up/Down difference (rms) & N/A & 0.007 &  0.011 \\
        Sum & 0.053 & 0.018 & 0.023 \\ \hline 
    \end{tabular}
    \caption{Systematic uncertainties and the estimated effect on the measured resonant frequency.}
    \label{tab:systematics}
\end{table*}

\section{Conclusions, Discussions and Future Work}
Understanding and quantifying ice properties such as the refractive index are critical inputs to polar UHEN experiments. 
Three different accuracy `scales' are relevant for neutrino measurements -- a) neutrino `identification' can be done by ignoring the RI variation with depth and using constant $n(z)$, since the primary parameter that separates radio signals produced by neutrino interactions from backgrounds is a simple up/down criterion - the former are upcoming, while backgrounds (primarily anthropogenic, but also including `natural' backgrounds such as those generated by the tribo-electric effect or down-coming cosmic rays) arrive from the surface, b) once identified, translating a statistically large sample of neutrinos into a flux requires an estimate of the neutrino `effective volume', defined as the equivalent ice volume over which neutrino interactions can be measured. For low detectable neutrino energies (10-100 PeV), the extent of the shadow zone dominates the estimated effective volume, with variations of 1\% in RIP translating into uncertainties in $V_{eff}$ an order-of-magnitude larger \citep{ali2024modeling}. Uncertainties in the $n(z)$ profile therefore directly translate into weaker upper limits on the neutrino flux at this energy scale. At `high' neutrino energies, $V_{eff}$ is determined by ice attenuation, so RI systematic uncertainties are less important. For neutrino `astronomy', for which signal reconstruction depends on precise ray tracing, 1\% refractive index uncertainties again translate into 10$\times$ larger uncertainties in arrival direction. 

In this paper, we have attempted to correlate antenna response with the index of refraction profile as a function of depth.
We find agreement between the shallow firn fluctuations implied by our IDP borehole $n(z)$ profile with direct density measurements, taken from the same borehole. 
Compared to other techniques that can be used to infer a refractive index profile, NPM (for example) is statistically superior to what can be achieved with antenna resonance. The advantage of the approach outlined here is speed (the 2025 data-taking required approximately 30 minutes of run time), does not require a winch or dedicated power source, uses hardware readily accessible to the UHEN experiments (antennas and a VNA), can be easily integrated into the RNO-G field plan and should be sensitive to RIP differences between two sites laterally separated by the 1--2 km typical spacing between UHEN receiver stations.

\subsection{Future Measurements}
Although UHEN experiments favor broadband antennas to maximize sensitivity to the impulsive, ns-scale signals produced by neutrinos interacting in-ice, refractive index measurements using this technique favor `ideal' (long, thin wire) narrow-bandwidth dipole antennas with high Q-values deployed into similarly narrow boreholes. The Stampfli drill operated by the Ice Drilling Program, for example, produces 57-mm diameter boreholes to a depth of 20 m, in three hours. Such a measurement would thereby minimize systematics due to air content in the hole as well as fitting uncertainties inherent in fitting $S_{11}(\omega)$ profiles. Even in such an ideal case, however, an irreducible systematic error due to the fact that one must sample over meter-scale wavelengths, and therefore average over smaller-scale fluctuations, remains.
In 2025 and 2026, the RNO-G experiment will systematically obtain additional $S_{11}$ from $\approx$10 newly-drilled holes at the Summit Station site, offering a considerably extended data sample for similar studies in the future.

\section*{Acknowledgements}
We thank the support staff at Summit Station for making RNO-G possible, as well as Jay Johnson and the UW-based Ice Drilling Program team for drilling the icehole from which the primary measurements in this paper were derived.  We are deeply indebted to the KU Physics and Astronomy Machine Shop, and particularly Scott Voigt and Mark Stockham, who designed and constructed the custom antenna used for recording the primary data essential to this measurement, as well as the spacers employed for antenna stabilization in 2025. We also acknowledge our colleagues from the British Antarctic Survey for building and operating the BigRAID drill for our project. Particular thanks are due Liz Morris and Bob Hawley for insightful conversations, and for availing us of their NPM data and field experience.

We would like to acknowledge our home institutions and funding agencies for supporting the RNO-G work; in particular the Belgian Funds for Scientific Research (FRS-FNRS and FWO) and the FWO programme for International Research Infrastructure (IRI), the National Science Foundation (NSF Award IDs 2112352, 2111232, 2111410, and collaborative awards 2310122 through 2310129), and the IceCube EPSCoR Initiative (Award ID 2019597), the Helmholtz Association, the Swedish Research Council (VR, Grant 2021-05449 and 2021-00158), the University of Chicago Research Computing Center, and the European Union under the European Unions Horizon 2020 research and innovation programme (grant agreements No 805486), as well as (ERC, Pro-RNO-G No 101115122 and NuRadioOpt No 101116890). 
\FloatBarrier

\bibliography{mybib}   

@article{barwick2005south,
  title={{South Polar in situ radio-frequency ice attenuation}},
  author={Barwick, S and Besson, Dave and Gorham, Pand and Saltzberg, D},
  journal={Journal of Glaciology},
  volume={51},
  number={173},
  pages={231--238},
  year={2005},
  publisher={Cambridge University Press}
}

@article{aguilar2022situ,
  title={{In situ, broadband measurement of the radio frequency attenuation length at Summit Station, Greenland}},
  author={Aguilar, JA and Allison, P and Beatty, JJ and Besson, D and Bishop, A and Botner, Olga and Bouma, S and Buitink, S and Cataldo, M and Clark, BA and others},
  journal={Journal of Glaciology},
  volume={68},
  number={272},
  pages={1234--1242},
  year={2022},
  publisher={Cambridge University Press}
}

@article{robin1969interpretation,
  title={Interpretation of radio echo sounding in polar ice sheets},
  author={Robin, G and Evans, S and Bailey, Jeremy T},
  journal={Philosophical Transactions of the Royal Society of London. Series A, Mathematical and Physical Sciences},
  volume={265},
  number={1166},
  pages={437--505},
  year={1969},
  publisher={The Royal Society London}
}

@inproceedings{fujita2000summary,
  title={A summary of the complex dielectric permittivity of ice in the megahertz range and its applications for radar sounding of polar ice sheets},
  author={Fujita, Shuji and Matsuoka, Takeshi and Ishida, Toshihiro and Matsuoka, Kenichi and Mae, Shinji},
  booktitle={Physics of ice core records},
  pages={185--212},
  year={2000},
  organization={Hokkaido University Press}
}

@article{prohira2021radar,
  title={{The Radar Echo Telescope for Cosmic Rays: Pathfinder experiment for a next-generation neutrino observatory}},
  author={Prohira, Steven and de Vries, KD and Allison, Patrick and Beatty, J and Besson, David and Connolly, Amy and Dasgupta, Paramita and Deaconu, Cosmin and De Kockere, S and Frikken, D and others},
  journal={Physical Review D},
  volume={104},
  number={10},
  pages={102006},
  year={2021},
  publisher={APS}
}

@article{Schr_der_2017,
   title={Radio detection of cosmic-ray air showers and high-energy neutrinos},
   volume={93},
   ISSN={0146-6410},
   url={http://dx.doi.org/10.1016/j.ppnp.2016.12.002},
   DOI={10.1016/j.ppnp.2016.12.002},
   journal={Progress in Particle and Nuclear Physics},
   publisher={Elsevier BV},
   author={Schröder, Frank G.},
   year={2017},
   month=mar, pages={1–68} }

@article{aguilar2023radiofrequency,
  title={{Radiofrequency ice dielectric measurements at Summit Station, Greenland}},
  author={Aguilar, Juan Antonio and Allison, Patrick and Besson, Dave and Bishop, Abby and Botner, Olga and Bouma, Sjoerd and Buitink, Stijn and Cataldo, Maddalena and Clark, Brian A and Couberly, Kenny and others},
  journal={Journal of glaciology},
  volume={69},
  number={278},
  pages={1929--1940},
  year={2023},
  publisher={Cambridge University Press}
}

@article{besson2023polarization,
  title={{Polarization angle dependence of vertically propagating radio-frequency signals in South Polar ice}},
  author={Besson, Dave Z and Kravchenko, Ilya and Nivedita, Krishna},
  journal={Astroparticle Physics},
  volume={144},
  pages={102766},
  year={2023},
  publisher={Elsevier}
}

@article{allison2019measurement,
  title={Measurement of the real dielectric permittivity $\varepsilon$r of glacial ice},
  author={Allison, P and Archambault, S and Auffenberg, J and Bard, R and Beatty, JJ and Beheler-Amass, M and Besson, DZ and Beydler, M and Chen, CC and Chen, CH and others},
  journal={Astroparticle Physics},
  volume={108},
  pages={63--73},
  year={2019},
  publisher={Elsevier}
}

@article{kravchenko2011radio,
  title={{Radio frequency birefringence in South Polar ice and implications for neutrino reconstruction}},
  author={Kravchenko, I and Besson, D and Ramos, Andres and Remmers, Juliet},
  journal={Astroparticle Physics},
  volume={34},
  number={10},
  pages={755--768},
  year={2011},
  publisher={Elsevier}
}

@article{allison2020long,
  title={Long-baseline horizontal radio-frequency transmission through polar ice},
  author={Allison, P and Archambault, S and Beatty, JJ and Besson, DZ and Chen, CC and Chen, CH and Chen, P and Christenson, A and Clark, BA and Clay, W and others},
  journal={Journal of Cosmology and Astroparticle Physics},
  volume={2020},
  number={12},
  pages={009},
  year={2020},
  publisher={IOP Publishing}
}

@article{Hawley_Morris_2006, title={Borehole optical stratigraphy and neutron-scattering density measurements at Summit, Greenland}, volume={52}, DOI={10.3189/172756506781828368}, number={179}, journal={Journal of Glaciology}, author={Hawley, Robert L. and Morris, Elizabeth M.}, year={2006}, pages={491–496}}

@article{RNOG_Philipp,
  author = "Windischhofer, Philipp",
  title={{Calibrating the Radio Neutrino Observatory in Greenland}},
  doi = "10.22323/1.470.0003",
  journal = "PoS",
  year = 2024,
  volume = "ARENA2024",
  pages = "003"
}

@article{morris2003density,
  title={Density measurements in ice boreholes using neutron scattering},
  author={Morris, Elizabeth M and Cooper, J David},
  journal={Journal of Glaciology},
  volume={49},
  number={167},
  pages={599--604},
  year={2003},
  publisher={Cambridge University Press}
}

@article{huege2017radio,
  title={Radio-wave detection of ultra-high-energy neutrinos and cosmic rays},
  author={Huege, Tim and Besson, Dave},
  journal={Progress of Theoretical and Experimental Physics},
  volume={2017},
  number={12},
  pages={12A106},
  year={2017},
  month={11},
  issn={2050-3911},
  publisher={Oxford University Press},
  doi = {10.1093/ptep/ptx009}
}

@article{drude1900elektronentheorie,
  title={Zur Elektronentheorie der Metalle},
  author={Drude, Paul},
  journal={Annalen der physik},
  volume={306},
  number={3},
  pages={566--613},
  year={1900},
  publisher={Wiley Online Library}
}

@book{lorentz1916theory,
  title={The theory of electrons and its applications to the phenomena of light and radiant heat},
  author={Lorentz, Hendrik Antoon},
  volume={29},
  year={1916},
  publisher={GE Stechert \& Company}
}

@article{ali2024modeling,
  title={Modeling the refractive index profile n (z) of polar ice for ultra-high energy neutrino experiments},
  author={Ali, S and Allison, P and Archambault, S and Beatty, JJ and Besson, DZ and Bishop, A and Chen, P and Chen, YC and Clark, BA and Clay, W and others},
  journal={arXiv preprint arXiv:2406.00857},
  year={2024}
}

@article{Askaryan1965CoherentRE,
  title={{Coherent Radio Emission from Cosmic Showers in Air and in Dense Media}},
  author={G. A. Askar’yan},
  journal={Journal of Experimental and Theoretical Physics},
  year={1965},
  url={https://api.semanticscholar.org/CorpusID:118128520},
   loc = {http://www.jetp.ras.ru/cgi-bin/dn/e_021_03_0658.pdf}
}

@article{Askaryan:1961,
    author = "Askar'yan, G. A.",
    title = "{Excess negative charge of an electron-photon shower and its coherent radio emission}",
    journal = "Zh. Eksp. Teor. Fiz.",
    volume = "41",
    pages = "616--618",
    year = "1961",
    url={https://api.semanticscholar.org/CorpusID:126197224}
}

@article{agarwal2025instrument,
  title={{Instrument design and performance of the first seven stations of RNO-G}},
  author={Agarwal, S and Aguilar, JA and Alden, N and Ali, S and Allison, P and Betts, M and Besson, D and Bishop, A and Botner, Olga and Bouma, S and others},
  journal={Journal of Instrumentation},
  volume={20},
  number={04},
  pages={P04015},
  year={2025},
  publisher={IoP Publishing}
}

@article{alden2025observation,
  title={Observation of In-ice Askaryan Radiation from High-Energy Cosmic Rays},
  author={Alden, N and Ali, S and Allison, P and Archambault, S and Beatty, JJ and Besson, DZ and Bishop, A and Chen, P and Chen, YC and Chen, Y-C and others},
  journal={arXiv preprint arXiv:2510.21104},
  year={2025}
}

@article{KOVACS1995245,
title = {The in-situ dielectric constant of polar firn revisited},
journal = {Cold Regions Science and Technology},
volume = {23},
number = {3},
pages = {245-256},
year = {1995},
issn = {0165-232X},
doi = {https://doi.org/10.1016/0165-232X(94)00016-Q},
url = {https://www.sciencedirect.com/science/article/pii/0165232X9400016Q},
author = {Austin Kovacs and Anthony J. Gow and Rexford M. Morey}
}

@article{vandecrux2020firn,
  title={The firn meltwater Retention Model Intercomparison Project (RetMIP): evaluation of nine firn models at four weather station sites on the Greenland ice sheet},
  author={Vandecrux, Baptiste and Mottram, Ruth and Langen, Peter L and Fausto, Robert S and Olesen, Martin and Stevens, C Max and Verjans, Vincent and Leeson, Amber and Ligtenberg, Stefan and Kuipers Munneke, Peter and others},
  journal={The Cryosphere},
  volume={14},
  number={11},
  pages={3785--3810},
  year={2020},
  publisher={Copernicus Publications G{\"o}ttingen, Germany}
}

@article{nghiem2012extreme,
  title={The extreme melt across the Greenland ice sheet in 2012},
  author={Nghiem, SV and Hall, DK and Mote, TL and Tedesco, Marco and Albert, MR and Keegan, K and Shuman, CA and DiGirolamo, NE and Neumann, G},
  journal={Geophysical Research Letters},
  volume={39},
  number={20},
  year={2012},
  publisher={Wiley Online Library}
}

@article{Herron1980FirnDA,
  title={{Firn Densification: An Empirical Model}},
  author={Michael M. Herron and Chester C. Langway},
  journal={Journal of Glaciology},
  year={1980},
  volume={25},
  pages={373 - 385},
    DOI={10.3189/S0022143000015239},
  url={https://api.semanticscholar.org/CorpusID:127224110}
}

@article{welling2024brief,
  title={{Brief communication: Precision measurement of the index of refraction of deep glacial ice at radio frequencies at Summit Station, Greenland}},
  author={Welling, Christoph and others},
  journal={The Cryosphere},
  volume={18},
  number={7},
  pages={3433--3437},
  year={2024},
  publisher={Copernicus Publications G{\"o}ttingen, Germany}
}

@Article{gmd-13-4355-2020,
AUTHOR = {Stevens, C. M. and Verjans, V. and Lundin, J. M. D. and Kahle, E. C. and Horlings, A. N. and Horlings, B. I. and Waddington, E. D.},
TITLE = {{The Community Firn Model (CFM) v1.0}},
JOURNAL = {Geoscientific Model Development},
VOLUME = {13},
YEAR = {2020},
NUMBER = {9},
PAGES = {4355--4377},
URL = {https://gmd.copernicus.org/articles/13/4355/2020/},
DOI = {10.5194/gmd-13-4355-2020}
}

@article{Salamatin_Lipenkov_Duval_1997, 
title={{Bubbly-ice densification in ice sheets: I. Theory}}, 
volume={43}, 
DOI={10.3189/S0022143000034961}, 
number={145}, 
journal={Journal of Glaciology}, 
author={Salamatin, Andrey N. and Lipenkov, Vladimir Ya and Duval, Paul}, 
year={1997}, 
pages={387–396}}

@article{schytt1958snow,
  title="{Snow Studies at Maudheim: Snow Studies Inland: the Inner Structure of the Ice Shelf at Maudheim as Shown by Core Drilling}",
  author={Schytt, Valter},
    ISSN = {16513215},
 URL = {http://www.jstor.org/stable/520189},
 journal = {Geografiska Annaler},
 number = {1},
 pages = {85--87},
 publisher = {[Wiley, Swedish Society for Anthropology and Geography]},
 reviewed-author = {V. Schytt},
 urldate = {2025-03-14},
 volume = {40},
doi = {https://doi.org/10.2307/520189},
 year = {1958}
}

@article{horhold2011densification,
  title={The densification of layered polar firn},
  author={H{\"o}rhold, MW and Kipfstuhl, Sepp and Wilhelms, Frank and Freitag, Johannes and Frenzel, Andreas},
  journal={Journal of Geophysical Research: Earth Surface},
  volume={116},
  number={F1},
  year={2011},
  publisher={Wiley Online Library}
}

@article{fausto2018snow,
  title={{A snow density dataset for improving surface boundary conditions in Greenland ice sheet firn modeling}},
  author={Fausto, Robert S and Box, Jason E and Vandecrux, Baptiste and Van As, Dirk and Steffen, Konrad and MacFerrin, Michael J and Machguth, Horst and Colgan, William and Koenig, Lora S and McGrath, Daniel and others},
  journal={Frontiers in Earth Science},
  volume={6},
  pages={51},
  year={2018},
  publisher={Frontiers Media SA}
}

@book{cuffey2010physics,
  title={The physics of glaciers},
  author={Cuffey, Kurt M and Paterson, William Stanley Bryce},
  year={2010},
  publisher={Academic Press}
}

@article{SDoniach_1970,
doi = {10.1088/0022-3719/3/2/010},
url = {https://dx.doi.org/10.1088/0022-3719/3/2/010},
year = {1970},
month = {feb},
publisher = {},
volume = {3},
number = {2},
pages = {285},
author = {S Doniach and M Sunjic},
title = {Many-electron singularity in X-ray photoemission and X-ray line spectra from metals},
journal = {Journal of Physics C: Solid State Physics}
}

@article{PhysRevD.98.030001,
  title = {Review of Particle Physics},
  author = {Tanabashi, M. and others},
  collaboration = {Particle Data Group},
  journal = {Phys. Rev. D},
  volume = {98},
  issue = {3},
  pages = {030001},
  numpages = {1898},
  year = {2018},
  month = {Aug},
  publisher = {American Physical Society},
  doi = {10.1103/PhysRevD.98.030001},
  url = {https://link.aps.org/doi/10.1103/PhysRevD.98.030001}
}

@article{Cowan_2019,
title = "Statistical models with uncertain error parameters",
author = "Glen Cowan",
year = "2019",
month = feb,
doi = "10.1140/epjc/s10052-019-6644-4",
language = "English",
volume = "79",
pages = "1--17",
journal = "European Physical Journal C",
issn = "1434-6044",
publisher = "Springer New York",
}
\bibliographystyle{igs}  

\end{document}